# Giga fine-structure constant in photonic solitons

Yao Cheng[1] and Ben-Li Young[2]

[1]Department of Engineering Physics, Tsinghua University, Beijing, Haidian, 100084, China

[2]Department of Electrophysics, National Chiao Tung University, Hsinchu, 300, Taiwan

**Abstract**

We report on the flavour physics of mesoscopic instantons found primarily in a pure Nb metal, doped magnetically by long-lived Mossbauer $^{93m}$Nb γs. A giga fine-structure constant $N_I\alpha$ characterizes various properties of the soliton, in which $N_I \sim 3\times10^{10}$ is the instanton number. A hierarchy of photonic solitons is classified into six flavours, i.e., three generations in each of two homotopy classes. The underdoped metallic heterostructure contains two topologically distinct sectors. When the γ-dopant filling factor ν > 0.88, the nontrivial sector becomes a chiral superfluid at $T \gg 300$ K to provide a γ-clock Q-factor > $10^{27}$ for some decades, while the trivial sector becomes a superconductor below 9.3 K. Distinct soliton flavours carrying Kopnin masses > 14 PeV, Majorana mass < 36 aeV, and nuclear $B$ > 35000 T imply a range of possible applications, e.g., topological quantum computation and the detection of invisible particles.

*1. Introduction*

The topological magnetoelectric effects (TME) [1] of a Dirac/Weyl metal [2, 3] embedded in a nonlinear, non-Hermitian, [4, 5] magneto-opto-mechanical, [6] photonic crystal [7-9] manifest themselves in a masquerade of indistinguishable particles involved in a game of gain-and-loss, including a long-lived Mossbauer $^{93m}$Nb photonic soliton [10-15] dressed by nuclear excitons [16] along with various high-order solitons [17]. The half-filling Nb metal becomes a heterostructure of topological quantum matter [18] after neutron bombardment. When the γ-filling factor ν is < 1, two topologically distinct sectors of the metallic and photonic phases are manipulated according to a number of parameters. The spin-4 Mossbauer γ introduces an emergent nontrivial Yang-Mills field for constructing an instanton [19, 20], which has a mass gap provided by alternate appearances of two distinct binding gauge fields locked in their common rotating reference frame, i.e., a $Z_2$-dimer time chain between the strong and electromagnetic (EM) interactions. The EM wave is then no longer simply an abelian gauge boson, rather it has become a non-abelian gauge boson that carries a spin charge to transform the soliton flavours, as does the gravitational wave (GW). An instanton number $N_I$ ~ 3×10$^{10}$ accounts for fractional fluxes in the p-wave soliton, which are localized at $\pm N_I$ charge zeros on the spacetime lattice sites. When ν > 0.88, the mesoscopic solitons enter a superfluidic phase at a temperature $T$ >> 300 K [12, 15].

Antiferromagnetic (AF) ladders of anisotropic instanton rods constitute a 2D magnetic lattice of vortices and antivortices in a field of zero $H$, and these restore the spontaneously broken time-reversal ($\mathcal{T}$), particle-hole (PH), and chiral symmetries. The chiral superfluid of AF ladders is like a Haldane Chern insulator [21] in the absence of Landau levels (LLs). A Floquet drive current [9] introduces next-nearest-neighbour hopping to mix the antiparallel and parallel ladders above an onset temperature of $T_{on1}$, which provides nonzero Chern numbers by breaking the $\mathcal{T}$, PH, and chiral symmetries spontaneously at the edge states, in a continuous manner with the number of parallel ladders being proportional to the drive current. We have observed two types of onset temperatures to turn the edge modes on, determined by whether or not the restored symmetries are broken when a gap is closed. The first type breaks the symmetries while reopening an active gap to maintain the edge zero modes in between. The second type simply switches the active gaps like quantum Hall effects, which provide a temperature-dependent Majorana mass [22]. The Chern insulator [21] implanted in the Weyl metal [2] has a physical dimension greater than three. We report on the Majoranization of AF solitons in the number-breaking phase as related to hypothetical neutrino physics [22]. The condensed spin-0 bosons at the zero-energy edge state acquire the emergent Majorana mass from the mHz hyperfine coupling [16] between the nuclear and soliton orbits.

Anisotropic solitons have a characteristic $N_I$ tag, i.e., emergent gauge fields of $\pm N_I$ charges, a giga fine-structure constant $N_I \alpha$, a Kopnin mass [23] of $N_I$ electrons, a non-abelian theta term [24] of $\theta_L = 2n_L \pi / N_I$, a chiral anomaly characterized by $N_I$, and several high-order chiral anomalies characterized by $N_I^2$. The applied $H$ field tunes an elementary winding number $n_L$ of the Hopf fibration to map the $S^3$ instanton to the $S^2$ monopoles [25]. The static and periodic drive fields cooperate to generate an $N_I$-gauge structure, which involves Higgs transitions, homotopy classes, quaternions, octonions [25], fractional electric/magnetic charges, anyons, and axions [18]. γ-ray scattering by instantons at ν >> 1 [15] demonstrates the extreme matter-field coupling of $N_I \alpha = N_I e^2 / 4\pi\epsilon_0 \hbar c$, i.e., the MeV γ ray turns immediately through 90° at the impinging surface caused by a photon blockade [9], where $\epsilon_0$ is the permittivity of free space, $e$ is the elementary charge, $c$ is the speed of light in a vacuum, and $\hbar = h/2\pi$ is the reduced Plank's constant.

Octonion solitons are unconventional, in that three-body interaction among octonions of $A$, $B$ and $C$ does not obey the associative law of $(AB)C = A(BC)$. Although the comprehensive non-associative physics is not yet fully understood, we herein demonstrate their associative memory. Just as it is very natural for elementary particles, atoms, molecules, and genes to remember information about their birth during their lifetimes, solitons also retain their birth identifications until they die. However, the twist memory is something rather different, and may sleep following a phase transition. The Goldstone zero mode of twisted instantons is woken up, when the proper morning call arrives. Note that the twists break the instanton $\mathcal{T}$-symmetry continuously rather than in a discrete manner.

For example, four parallel ladders constitute a gyrating octonion at 300 K, which changes to two quaternions in the superconducting state when $T$ is lowered below a critical $T_c$. The twist memory of gene-like double spirals is preserved when passing $T_c$, even at very small $H$. Hidden edge states emerge, when the proper drive is applied. We frequently observed the hysteresis regarding $T$, $H$, ν, and *etc*., yielding different twists by driving the parameters up and down passing some critical points in a thermal equilibrium. AF ladders of [93m]Nb instantons in TME experiments remember a commensurate twist provided by the $H$ field and the drive frequency $f_d$ when entering the superconducting phase. AF ladders of [103m]Rh instantons in x-ray experiments remember the twists provided by ν at the moment when pumping stops at 300 K.

We herein provide a long list of conditions that validate the exceptional fine-structure constant $N_I \alpha$ for topological actions. The first involves hyper Raman scattering of MeV γ rays [15], in which the extremely small nuclear cross section of nonlinear four-wave mixing is amplified by $N_I^4$ [9]. Scattering of randomly impinging MeV γs for 107 days [15] showed seven Sisyphus cycles in one of the end-fire modes [6], which propagated unidirectionally at the interface between a chiral superfluid and a vacuum. The scattering experiments revealed broken $\mathcal{T}$-symmetry, broken parity ($\mathcal{P}$) symmetry, pronounced Kerr nonlinearity, and a topological γ-laser [9]. Under a low-frequency Floquet drive of several Hz, the TME gains energy from time to time. Hence, the $\mathcal{PT}$-symmetry of the Floquet system may also be broken, as a result of the coupling of the non-Hermitian optical system with a nontrivial vacuum.

The entangled multiple γs [12, 14] (see also Fig. A2 in Appendix A) acquire a mass by absorbing the longitudinal phonon of resonant nuclei [18] such that the metallic crystal becomes photonic below a critical temperature $T$ corresponding to a Rabi gap $\hbar\omega_R$ ~ 84 eV of $^{93m}$Nb, as revealed by the Mollow triplet [26] in the γ spectrum [10,14]. Strong coupling between γ and $^{93}$Nb atoms causes the photonic mass gap [26]. The Yukawa γ tail ~ 10 nm isolates the anisotropic instanton rods in spatial terms. The decay of delocalized $^{93m}$Nb is no longer a constant, instead it depends on $\nu$, $T$ [12], *etc*. The instanton is a zero-mode Kondo soliton that floats energetically on the Fermi surface [27], which collectively locks or unlocks the orbits of $N_I$ spinless electrons to screen $N_I$ polarized nuclear magnets when crossing the Kondo temperatures, e.g., $T_{Kl}$ ~ 56 K and $T_{Ka}$ ~ 61 K. Ions and electrons in the Nb crystal live in two different worlds, distinguished by the freedom-crossing wall.

Instantons are elementary building blocks used to construct a hierarchy of solitons. Solitons are generally classified into two topological types with distinct Euler characteristics [25]. Solitons of the first type align themselves in wormhole-like chains that penetrate the sample whilst being pinned by two boojums [23] at the sample boundary. Parallel ladders gyrate to create twists between the two boojums, which provide the Goldstone zero modes. Solitons of the second type loop themselves as movable vortons [23] free from boojums. Some twists may persist for a while, and parallel ladders detach themselves from the boojums to construct vortons. Both types of soliton contain three hierarchical generations of distinct homotopy classes, e.g., $S^3$, $S^4$, and $S^8$ solitons. Hence, we use the term 'flavours' to distinguish six major species of solitons in analogy with the standard model. The Lifshitz transition [28] between the flavours shows hysteresis depending on $\nu$, $T$, $H$, $f_d$, and the operational procedures, where distinct flavours may preserve their birth twists as intrinsic properties.

As for the Dirac quantization states [24, 25], whenever a charge $e$ is present, the monopole magnetic charge $g$ must be quantized by $eg = 2n\pi\hbar$ and vice versa, which is equivalent to the $2n\pi$ flux quantization of the 2D magnetic lattice with the unit $\hbar/e$. The magnetic soliton has a spontaneous $2\pi$ flux with a minimal nonzero integer $n = 1$. For the Onsager-Feynman quantization [29], it is further stated that the quantization of the circulating particle is inversely proportional to its mass. Accordingly, the ionic flux $2n\pi/N_I$ on the crystal lattice sites is fractional, which violates the Dirac quantization. As a remedy, carriers must have $\pm N_I$ charges bound together as a whole, which is the central consideration in the present report. It seems that there are two gauges inside the instanton, namely $e_1 = e$, $g_1 = 2n\pi\hbar/e$ and $e_2 = N_I e$, $g_2 = 2n\pi\hbar/N_I e$. The Dirac-Zwanziger condition [24, 30] $e_1 g_2 - e_2 g_1 = 2n\pi\hbar$ prevents arbitrary motion of $e_1$ and $e_2$, as revealed by ions and Mott electrons in the nontrivial sector, and the itinerant electrons in the trivial sector.

An instanton has the appearance of an extremely heavy particle, in which the four masses of the electron, ion, proton, and nucleus play predetermined roles. In precise terms, the measured value of $N_I$ is

$$N_I = \left(\frac{M_{Nb}}{M_e}\right)^2 \cong 2.86817 \times 10^{10}, \quad (1)$$

this being the quadratic ratio between the Nb$^+$ ionic mass $M_{Nb}$ and the electron mass $M_e$. The square power reflects the deep origin of the Berry curvature $\Omega_B$ of the mass dimension $[M]^2$ [31]. To see this, we take opposite $\pm N_I$ charges of ions and electrons bound together with two Kopnin masses of $N_I M_{Nb}$ and $N_I M_e$. They slosh around the centre of mass (COM) with a reduced mass of $N_I M_e M_{Nb}/(M_e + M_{Nb}) \approx N_I M_e$, as shown in Fig. 1a, which is 14.6564 PeV. Ions and electrons see different $B_\pm$ but the same Berry curvatures $\Omega_B$, where the symbols + and – apply to ions and electrons, respectively. We now introduce in-plane coherent $\dot{r}_\pm$ motions on the horizontal plane, which characterize the relatively slow $\omega_R$ described in the following paragraphs, while leaving the zitterbewegung that characterize the fast $\omega_\gamma$ of 30-keV $^{93m}$Nb until the next section. The decomposable dimensionality will become clear later on.

We define semi-classically $\dot{r}_\pm = \partial \varepsilon_{p\pm}/\partial p_\pm + \dot{p}_\pm \times \Omega_B$ and $\dot{p}_\pm = \pm eE \pm (e/c)\dot{r}_\pm \times B_\pm$ without LLs [32], where $r_\pm$, $p_\pm$ are the coordinates and the quasi-momenta, and $\varepsilon_{p\pm}$ are the quasiparticle energies. The former equations carry no charge, whereas the latter carry an electric charge. The gauge-invariant $\Omega_B$ offers a topological equivalence principle of the pseudo-spins under a magnetic field [25] regardless of the interaction types. Particles of distinct masses are free-falling in the rotating reference frame. QED and QCD work with a common $\Omega_B$ to maintain a nuclear version of γ clocks [33] characterized by a Q factor > $10^{27}$, which is important in the fundamental physics [34].

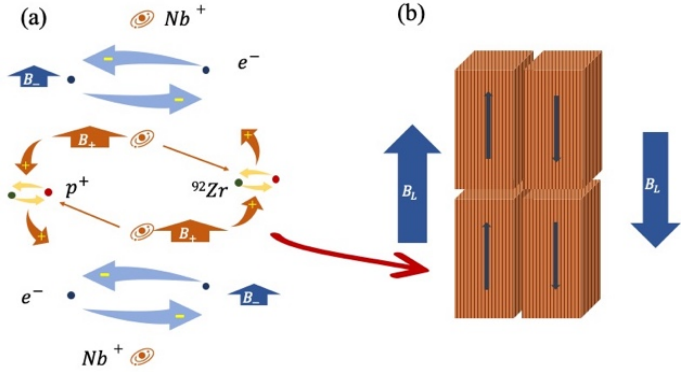

Figure 1, a) Instanton dimer chain on crystal lattice sites. Two quantum circles in the dimer are labelled by + and – signs, which consist of electrons $e^-$, ions Nb$^+$, protons $p^+$, and $^{92}$Zr nuclei under $B_\pm$ with a common static instanton flux and opposite charge flows. The fm nuclear circulation is exaggerated in the figure. b) Mesoscopic AF magnetic lattice. We simplify the hexagon-like rods of instantons using square rods, which are $n_A = N_I/n_L$ dimer chains stacking over $n_L$ layers. The flux density $B_L$ now accounts for $n_A$ fractional fluxes over the cross section.

The persistent multipolar flopping between nuclei and atoms gives no flux contribution to the static fluxes inside the pure-gauge instanton. Two magnetic lengths [35] $l_\pm = \sqrt{\hbar/n_A e B_\pm}$ define the mass ranges of $\pm N_I$ charges, which harbour $n_A = N_I/n_L$ fractional fluxes in their own rotating reference frames (see Fig. 1a). This particular COM condition implies two distinct radii $M_{Nb}r_+^2 = M_e r_-^2$ encircling two distinct fields $M_e B_+ = M_{Nb} B_-$. More precisely, linear momenta vanish not in one layer but separately in four layers (and not in a single layer) with two COMs, with electron-to-electron and ion-to-ion circling in paired layers (Fig. 1a). A Peierls transition [18] dimerizes $N_I$ Nb atoms to obtain the charge- and pseudospin-density waves below a critical temperature $T_P \sim$ 3000 K above the melting point of Nb. The topological Peierls gap is also a Zeeman splitting of nuclear pseudospin under $B_+$, where $n_L$ chiral electron skipping occurs at the edge states (Fig. 1b). One edge electron per layer reveals the tunable Hopf winding number $n_L$ to be a function of the aspect ratio.

The Higgs mechanism breaks the SO(3) symmetry down to a U(1) symmetry to give the finite-energy solution of 't Hooft–Polyakov monopoles [17, 25, 36, 37]. We have topological monopoles with fractional magnetic charges $\pm h/n_A e$ in the **k**-space. A collection of $n_A$ fractional charges gives opposite magnetic charges $\pm h/e$. We have three kinds of topological monopoles, i.e., two from opposite charge carriers and one from instantons. The opposite charge flows of ions and electrons pair their monopoles to zero below $T_{Kl}$, leaving the instanton monopoles confined by a very weak force proportional to $n_L$, as revealed by the aspect ratio ~ 4 of the instanton rods. If the confinement force is strong, we have a nearly isotropic instanton rod. The Dirac monopoles at the boojums are assisted by instanton fluxes in the **r**-space, which cancel each other either in AF ladders or in vortons. Static fractional fluxes of periodic $B_\pm$ distributions thread though $n_L$ layers of $r_\pm$ quantum circles in $n_A$ $Z_2$-dimer chains, which cause the electric and magnetic polarizations in the $S^3$ soliton. A Peierls gap $\Delta_P \sim 0.3$ eV protects the coherent circulation $\dot{r}_+$ of ions nailed down on lattice sites, whereas a Kondo minigap [23] ~ 5 meV protects the coherent circulation $\dot{r}_-$ of sprightly Mott electrons.

AF ladders are an internal γ field in the nontrivial sector, which is gapped by a superconductor in the trivial sector below $T_c$. We apply an external γ as the Floquet drive [38] to create a macroscopic midgap state of the Majorana zero mode (MZM) [22], which is localized in place of a gap-closing quantum phase transition [2, 9, 18]. The MZM remains real and acquires no TME phase advance, when the external and internal γ fields are exactly commensurate. This Majoranization is useful for identifying invisible cosmic fields, which play the role of an external drive. Accordingly, we have several distinct MZM flavours, as elaborated later on. However, a near-commensurate γ drive may split the vanishing monopoles of matter fields and may activate the complex Dirac fields on a complex base space [25]. The internal kinetics of ions and electrons give rise to a nonzero TME phase, as long as they escape the Majorana condition.

Of particular interest in the present report is a Floquet time crystal [39], e.g., an anomalous Floquet Chern insulator at zero net $H$ [9]. The Floquet γ drive (see Fig. 1c and 1d) introduces an extra time dimension to excite the instantons into two higher generations. For example, an $S^4$ soliton in the

(3+2)D heterostructure is the $S^4$ coboundary of a 5D ball along two time axes [24, 25]. One fast time scale refers to the hadron dynamics while one slow time scale refers to the soliton dynamics. According to the Derrick's theorem [17], any soliton will be located on a stationary point at the energy minimum. The nonlinear dynamics further demand a commensurate condition between two axes of time to protect the integrability by a winding number. These two guidelines are crucial for describing the solitons, as presented below.

We focus firstly on the $S^4$ solitons. The Floquet drive breaks the $\mathcal{T}$-symmetry spontaneously and then excites a two-level superposition continuously in the superfluidic bulk, i.e., antiparallel ladders on the groundstate and parallel ladders on the lowest LL (LLL) excited state [35]. We must be cautious in applying the naive concept that parallel vortices gyrate [31] because of the hexagon-like shape of the instanton chains pinned by boojums, which do not have as much freedom to gyrate as free-moving particles. The unidirectional phase advance of LLL "gyration" thus refers to the phase-coherent precession of canted orbital antiferromagnets in the Bose-Einstein condensation (BEC) of magnons [40], which is the condensed pseudospin-density wave.

The (3+1)D chiral edge states of the broken PH-, $\mathcal{T}$- and chiral symmetries are separated by a domain wall [18] where the Majorana mass changes sign [22]. According to the equivalence principle, the instanton condensate splits into dual macroscopic edge states, which are free-falling in their own rotating reference frames. The Dirac nodes have two split side bands of Weyl nodes as a Kondo triplet, which are split by a topological cyclotron gap $hn_H f_{LLL}$, i.e., the hyperfine Majorana mass gap [22]. More precisely, the amplitude of the side bands yields the LLL excitation number in continuous two-flavour mixing, which is proportional to the drive current. The internal and external periodic conditions are locked by $n_H$ in the (3+2)D crypto-equilibrium [39], which will be identified later on as the Hopf invariant of the second and third generations, to distinguish them from the Hopf invariant $n_L$ of the first generation.

We approach the hadron dynamics with $^{92}$Zr and a proton in $^{93m}$Nb$^+$ (Fig. 1a), where the EM interaction between an electron and the neutral $^{92}$Zr atom at the $0^+$ groundstate is negligible. An integer $\omega_\gamma/\omega_R \cong 4 \times 92$ is attributed to the $^{92}$Zr and the proton bound by the strong force. The number 92 comes from the Onsager-Feynman quantization, while the number 4 caused by the spin-4 quanta will be explained in the next section. Usually, the Rabi gap is proportional to the field-matter coupling strength [26]. In contrast, the mass gap $\hbar\omega_R$ of the γ Mollow triplet is determined by the hadron and lepton dynamics. A particular $f_d$ matches $f_{LLL}$ with $m_H f_d = 4n_H \times 92 f_{LLL}$, such that two triplets are commensurate with $m_H$ and $n_H$ coprime to form high-order solitons. Hadrons bound by the SU(3) QCD and ions/electrons bound by the M4 [12] QED share a common 2D projection of their stereoscopic dynamics (see Fig. 1a).

We find that the $Z_2$-LLL and $Z_4$-LLL vortices of the first topological type are attributed to the Hopf fibrations of two and four knotted quaternions $q$s in the Floquet time crystal, respectively.

The semionic instanton is the $S^3$ fibre of the unit quaternion $|q|=1$ in the Hopf map $\pi: S^7 \to S^4$, but there is no counterpart for the $S^7$ fibre in $\pi: S^{15} \to S^8$ as has been suggested by the physics, as highlighted in [25]. Here, the unit $Z_2$-LLL vortex of $|q^0|^2 + |q^1|^2 = 1$ is the $S^7$ fibre in $\pi: S^{15} \to S^8$, as a point-like $S^8$ soliton carrying the Cayley number [25]. Two winding numbers $n_H$ of the second and third generations correspond to these two Hopf fibrations. Flipping the $S^4$ and $S^8$ solitons emit GWs collectively, it is hardly surprising that their MZM quadrupoles receive GWs collectively as well. Four distinct gauge bosons match each other well to acquire a Majorana mass gap for the spin-3/2 MZM in a quantum manner, i.e., one external and two internal quanta invite a graviton from one and only one GW source to synchronize their rotating actions far away. Observations in this report reveal the Mach principle that all the distant stars compete in creating the Majorana mass.

We recognise further the $Z_2$-LLL vortex carrying a Grassmann number [22, 25] as the Moore-Read Pfaffian (MRP) state [35, 41], while the $Z_4$-LLL vortex corresponds to the Read-Rezayi parafemionic (RRP) state [35, 42]. Although $\pm N_I e$ charges bound to one $2\pi$ flux look like an abelian integer quantum Hall state carrying the first Chern number $C_1 = N_I$, this picture is in fact completely wrong. The MRP and RRP vortices are characterized by high-order Chern numbers $C_n$, being somewhat like the non-abelian quantum Hall states but in a different context. More precisely, anyon flux numbers in a 2D magnetic lattice constitute the second Chern number $C_2$ and the third Chern number $C_3$ [17, 24, 25] regarding the elementary and high-order chiral anomalies, respectively, while the first Chern number $C_1$ vanishes. The quaternions are non-commutative in favour of the topological quantum computation [43]. The octonions are neither commutative nor associative, and this could introduce an interesting representation for three generations of the standard model [44].

We also find a Lifshitz transition between two topological types, i.e., the $Z_4$-vortons emerge spontaneously by lowering $H_z$ above $T_{Ka}$, the flavour of which is transformed from $S^8$ to $\mathbb{T}^4 \times S^4$, while we can only speculate a similar means of transforming the $Z_2$-vortices from $S^4$ to $\mathbb{T}^2 \times S^2$ in cases where $T_{Ka} < T_c$. These are free-moving anapoles [45] of the second topological type grouped by four or two parallel instanton rings, where the magnetic monopoles vanish while the electric monopole-lines loop themselves in circles. A $Z_4$-vorton train constitutes the spacetime 4-torus of $\mathbb{T}^4$ under the periodic boundary condition, as shown later on. We hereafter refer to these as the gotwo and gofour generations of the anapole, to distinguish them from the standalone anapole $\mathbb{T}^2 \times S^1$ of the [103mRh] instanton, which emerges spontaneously when $T$ is lowered. The continuous transition caused by raising $T$ to break the $\mathcal{T}$-symmetry implies a subtle flavour physics to recover the hidden zero modes from twist memories. We also report on how the seesaw quadrupole (see Fig. 1d) couples the spin-2 field and how the gofour anapoles couple a spin-0 field (see Fig. B3 in Appendix B). In summary, six major flavours arise naturally from the Hopf maps, classified by opened and closed strings, i.e., the availability of boojums.

When doping electrons by reducing $v < 1$, a Fermi moat develops around AF-ladder wormholes to provide a Dirac metal at zero $H$. The metallic heterostructure is somewhat related to a well known superlattice model of the Dirac/Weyl metal as suggested by Burkov and Balents [46], which has disconnected Fermi sheets at the interfaces between the magnetically doped topological insulator layers and the normal insulator spacers [2]. We highlight here a number of differences between the two heterostructures. The most profound difference is related to the topological degrees of freedom, which allow the spin-4 condensate to couple with various external fields. The chiral anomaly fuses two spin-1 fields in a triangle diagram to form a spin-0 chiral gauge field [2], which is characterized by $N_I$ in our case. The high-order chiral anomalies either fuse four spin-1 fields in a non-abelian pentagon loop diagram [31], mix one spin-0 and four spin-1 fields in a hexagon diagram, or mix two spin-1 and one spin-2 fields in a square diagram, which are then characterized by $N_I^2 \sim 10^{21}$. The non-abelian $\theta$ terms arise from $C_2$ and $C_3$ rather than $C_1$ [24]. Hence, the quantum anomalous plateaux [2] are caused by axion electrodynamics [1] rather than 2D Hall transport [35]. The Fermi moat is connected arcwise, rather than simply [25]. There are two kinds of edge electron in the presence of Weyl nodes. The first kind appears at the interface between the wormholes and Fermi moat, and is neglected in the superlattice model. The second kind appears at boojums, and corresponds to the Weyl electrons in the superlattice model. When the underdoped $v$ exceeds a quantum-critical density [47] $v_{qc} \sim 0.88$, the instantons undergo BEC at $T \gg 300$K, as revealed by γ-ray scattering [15]. The nontrivial γ-superfluid and the trivial s-wave superconductor [48] coexist below a superconducting transition temperature $T_c$ of 9.3 K.

A sufficiently strong $H$ converts the Dirac metal to a Weyl metal. For example, an $H$ of 3-Oe order can only reorientate the AF ladders in a square sample sheet by 90° to maintain the superfluid coherency at room temperature, as revealed by monitoring the x- and γ-ray intensities of $^{103m}$Rh (see Fig. A1 in Appendix A). Creation of Weyl nodes in a superconducting disk at 9 K in the $^{93m}$Nb case requires a much stronger vertical $H_z > 500$ Oe (see Table 1). The surface superconductivity preserves the Dirac nodes, where the PH-Pfaffian (PHP) flavour of pairing composite fermions emerges under the global $\pi$ Berry phase [35, 49]. The PHP flavour contains an electronic gyration perpendicular to the AF ladders as characterized by the non-vanishing $C_1$, which is quite different from the vanishing $C_1$ in the MRP and RRP flavours.

A 6-Hz Floquet current $I_x(t)$ opens the domain wall spontaneously [18] along the central line of the superconductor by raising $T > T_{on1} \sim 8.1$ K near $v_{qc}$ at zero net $H$, which yields a macroscopic MZM by splitting the (3+2)D composite condensate into two cat states with continuous opposite magnetizations proportional to $I_x(t)$ (see Figures 1c, 1d, and B1 in Appendix B), i.e., MZM characterized by opposite Chern numbers [31]. $I_x$ closes the superconducting gap to break the $\mathcal{T}$-, PH-, and chiral symmetries while it reopens a dynamical gap to accommodate MZM. The thermal potential of $T_{on1}$ is necessary to release MZM in the s-wave superconductor. The particular $f_d = 6$ Hz is required to approach the near-commensurate condition of $|m_H f_d - 4n_H \times 92 f_{LLL}| < 1/\tau_{R1}$, where $m_H = 1$, $n_H = 1$, $f_{LLL} \sim 15.625$ mHz ± 8 μHz and $\tau_{R1} \sim 2$ seconds. The dipole relaxation time $\tau_{R1}$ is a characteristic of the nontrivial sector, which is nearly independent of $T < T_c$. The near-commensurate Floquet drive using $f_d = 6$ Hz has twist memory when lowering $T < T_c$ at the earth's field, which is woken up by raising $T > T_{on1}$. It is a pure quantum effect that the particular $f_d$ breaks the $\mathcal{T}$, PH, and chiral symmetries spontaneously in the weak Floquet $H$ limit.

According to the bulk-edge correspondence, AF ladders in the domain wall are transferred to clockwise and counterclockwise one-way gyrating MRP vortices. The diamagnetic mirrors of composite condensates confine the dynamic $H$ field to the domain wall, where $I_x(t)$ charges the boojums to provide an electric field $E_y(t)$. We measured the transverse terms along the y axis, which reveal clearly $E_y(t)$ from the chiral anomaly (see Fig. B2 in Appendix B). The field gradient drives the MRP ±π fluxes bouncing back and forth within the domain wall, as their own antiparticles but in different states of motion.

A quantum anomalous $\mathcal{T}$-plateau $V_y(t) = I_x(t)C_2 R_K/N_I$ emerges in an interval of $T_{on1} < T < T_c$, where $I_x(t)$ represents the anomalous Hall current, $C_2 = 2$, and $R_K = h/e^2 \approx 25.8$ kΩ is the von Klitzing constant [50]. The plateau is flat regarding the parameter space spanned by $f_d$, $T$, and $v$ in the weak $H$ limit, which reveals an incompressible quantum liquid, i.e., conserving two distinct flavours in a stroboscopic sense [39].

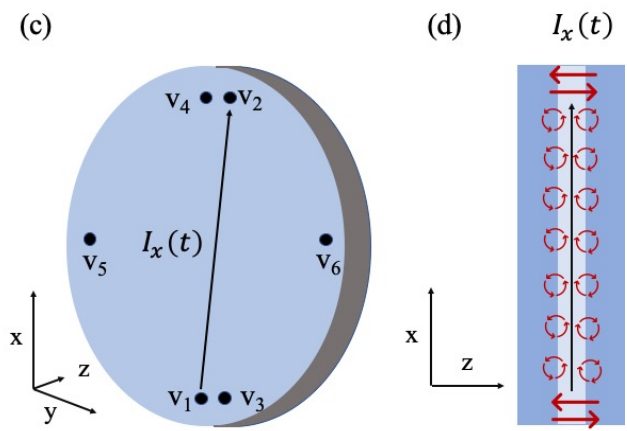

Figure 1, c) Oval sample disk of 13mm×12mm×1.2mm. Six via holes are indicated by $v_1$ to $v_6$, which feed the drive current between $v_1$ and $v_2$ and pick up the voltages between $v_3/v_5$ and $v_4/v_6$. The line $v_1$-$v_2$ crosses the line $v_3$-$v_4$ in the first test. They are nearly parallel in the second test, i.e., the positions are exchanged between $v_1$ and $v_3$. d) Condensates (blue) split by a domain wall (white), where the applied current $I_x(t)$ flows on an x-y plane at z = 0. The clockwise and counterclockwise MRP/RRP vortices along the relatively long y-axis in red are bouncing within the domain wall. The opposite magnetic and electric polarizations at ±z sides constitute the macroscopic MZM and a seesaw quadrupole $Q_y$, and their amplitude is modulated by the drive current.

A similar configuration in a $\mathcal{T}$-breaking $^3$He-A Weyl superfluid [51,52] has been investigated in the context of the Andreev-Majorana bound state, which is its own antiparticle. In our case, the macroscopic MZM emerges between two cat states of opposite magnetic polarizations. The composite condensate is either a Dirac or a Weyl superconductor depending on whether or not $H$ splits the Dirac nodes [53]. The latter topological superconductor is also related to the Weyl superconductor model as suggested by Meng and Balents [54]. The Floquet drive is necessary to obtain the nontrivial TME, otherwise the "relativistic" effect of the massless fermions vanishes, which leaves a trivial response in the thermal dynamic equilibrium [2].

We assume that the interface energy between two sectors vanishes at $T \sim 4.6$ K, as explained later on, near which the energy of the Bogomolny-Prasad-Sommerfield (BPS) solution of arbitrary non-abelian monopole numbers is bounded from below [17]. There is no force between the BPS MRP vortices. A Lifshitz transition between two flavours gives rise to two different TMEs as distinguished by hysteresis paths across a critical field $H_{c1}$ at 4.2 K, i.e., either a zero-TME superconductor in the $H_z$-up phase or the BPS MRP $H_z^2$-supermagnets, which preserve the octonion twists in the $H_z$-down phase.

The Chern-Simons 5-form $dS^4$ of the MRP zero modes is closed and locally exact according to Poincaré's lemma [25], where the annular magnetic drive field expelled to the superconductor boundary cannot change the number of MRP vortices. This $dS^4$ 5-form is thus characterized by the third Chern number $C_3$, where the complex tensor fields of charge flows provide the electric and magnetic polarizations. The complex base space of the torsion-free MRP solitons refers to the Calabi-Yau manifold with vanishing first Chern class [25]. These remarkable integrable dynamics are as important as the integer quantum Hall effect [50]. A $4R_K/N_I$ PHP quantum plateau appeared in the presence of Dirac nodes at nonzero $H_z$, whereas the $H_z^2$-plateau arose from two cascade topological pumps [9] in the presence of Weyl nodes.

Because end-to-end interaction is stronger than side-by-side interaction, the instanton rods align themselves along the longest x-axis of the oval x-y-disk sample (Fig. 1c), as revealed by x- and γ-rays (see Fig. A1 in Appendix A) [14,15]. The overlapping Yukawa tails force the instanton BEC at $v > v_{qc}$, which depends on the sample shape. More precisely, this is a Berezinskii-Kosterlitz-Thouless transition [55] of vortices and antivortices in the reducible dimensionality.

The SU(2) part of the multipolar gluon field provides eight Goldstone modes of the instanton [24]. More Goldstone modes arise in the condensate, e.g., the phase-coherent precession of the cat state [40] and a breathing mode of anapoles provided by the Floquet drive in the weak $H_z$ limit. Several Goldstone modes emerge in the presence of $H_z$, e.g., a swing-twist-buckle (STB) mode of RRP vortices pinned by boojums. Various zero modes span a kernel volume in the moduli space [17] of soliton configurations, where the twist is memorized in the zero-mode freedom of matter fields. The metallic heterostructure contains topological orders that depend not only on the parameter settings but also on the paths in arriving at these settings. Hence, the TME response may be different even when identical parameters are used.

We also investigated two more long-lived Mossbauer γ-dopants, i.e., $^{45m}$Sc and $^{103m}$Rh. These have unique characteristics of their own, i.e., long lifetime, multiple spin quanta, odd parity, metallic crystal, and most importantly identical isotopes of 100% natural abundance. The M4 γ-dopant of 30-keV $^{93m}$Nb has 4 spin quanta and a 16-year half-life, the E3 γ-dopant of 40-keV $^{103m}$Rh has 3 spin quanta and a 1-hour half-life, while the M2 γ-dopant of 12-keV $^{45m}$Sc has 2 spin quanta and a 0.3-second half-life. This report focuses on the extra long-lived $^{93m}$Nb particularly at ν < 1, which allows us to measure TME at low $T$ and high $H$. We focus also on the x-ray experiments of $^{103m}$Rh, which may have solitons with strong orbital mixing rather than with the p-wave solitons of $^{93m}$Nb.

The complexity of the science involved means that many of the experimental results described here may seem inconsistent; essentially, this is attributable to the path-dependent memory. Further details are therefore necessary to describe the operational sequences properly. Moreover, some signals are related to cosmic events, which also requires details of the dates of the experiments. Although many of the experimental results are not yet fully understood, rapid developments in our understanding of topological quantum matter [18] provide some help in exploring the complexities involved, many of which are documented in celebrated textbooks and review articles, and the references therein.

*1.1 The collective instanton*

Many kinds of groundstate degeneracy are available. For example, the extremely slow transition of four entangled γγγγ > $10^{17}$ years is amplified collectively by degenerate virtual emissions [14], which are launched in a time < 0.1 fs. The method of accounting for the enhancement is refined after acknowledging the instanton. $N_I^4$ degeneracy is proposed for joining up $N_I$ nuclei, $N_I$ phonons, and $N_I^2$ magnons in place of the previous proposal, and this accounts for the $\sqrt{N^4}$-fold superradiance of $N \sim 10^{22}$ nuclei in a crystal [14].

The chiral spin liquid [56-58] shows many degenerate configurations of instanton dimers inside a $\mathcal{T}$-breaking $\theta_L$ vacuum (see Fig. 1a), which provides another kind of enormous groundstate degeneracy. A similar groundstate degeneracy can be further extended to the triangle-like lattice of AF ladders, as revealed by the six-fold symmetry of the γ-ray scattering [15]. The instanton squeezes to a hexagon-like shape due to its high packing factor of $\nu > 0.88$ (see Fig. 1b). Hereinafter, we use the term "lotton" for the collective instanton, while referring to the instanton as the fractional instanton on crystal lattice sites.

Two breakthroughs in our efforts to explain the photonic soliton are the nuclear Kohn mode (NKM) and the quantum time crystal (QTC) [59, 60]. The NKM reveals the spatial textures of the $S^3$ soliton, while the QTC reveals the temporal structure. Under the celebrated Kohn theorem [61], the highly degenerate electron-hole pairs gyrating around COM slosh collectively as a whole, regardless of the details of the other interactions. The origin of the Kohn theorem arises from the forbidden

spin-0 transition to excite the untouchable Kohn mode without any momentum transfer, which is amplified collectively by the degeneracy. The NKM does this too, where the extremely forbidden transition of γγγγ is amplified collectively to provide a tough Rabi gap. NMK ions replace the holes in the conventional Kohn mode to break the PH symmetry. Every $^{93}$Nb atom in NKM is decomposed into one electron, one proton and one $^{92}$Zr atom, all of which gyrate coherently around their COM (Fig. 1a). It was Wilczek who introduced the QTC to describe a persistent oscillation in a superconductor. Most recently, several versions of the Floquet time crystal were found theoretically and experimentally, as revealed by the subharmonics of the periodic drive [39]. Our integrable soliton with the long-range multispin interaction [62] goes back to the original QTC concept of Wilczek [59, 60], where the quantum walk yields a subharmonic of $\omega_R \ll \omega_\gamma$.

The Nb atom has only one 5s electron outside the full 4d$_{3/2}$ shell, which becomes the Mott electron that causes the Rabi oscillation to persist between the sp- and pg-hybridizations at $T < T_{Kl}$. The spin-4 γ excitation of $^{93m}$Nb is provided by a nuclear transition between the groundstate $J^\pi = 9/2^+$ and the low-lying excitation $J^\pi = 1/2^-$ [63]. The groundstate has one proton $p^+$ on the outer shell of $1g_{9/2}$. The transition constitutes a proton jump from the underlying shell $2p_{1/2}$ to the pair $p^+$ on $1g_{9/2}$ while leaving the single $p^+$ on $2p_{1/2}$. We thus approximate the COM $^{93}$Nb$^+$ ion with the co-gyrating proton and the $^{92}$Zr atom is bound by the strong force (Fig. 1a). The Peierls transition moves two adjacent nuclei closer together as an instanton dimer, where the spin-4 γ becomes a generalized resonating valence bond [64]. The $\omega_\gamma$-zitterbewegung of protons/nuclei in instanton dimers provides coaxially an M2 magnetic quadrupole and an E2 electric quadrupole to represent the time-varying M4 γ field [63].

The QTC proceeds via a magnetic quantum walk that strictly follows a $Z_4$ periodic protocol. $\omega_\gamma = 4 \times 92\omega_R$ guarantees the pronounced quarter γ ray as one of the γγγγ [14], otherwise arbitrary γ rays of energy less than quarter γ appear to consume the energy, the composite wavelengths of which can no longer match the lattice constants. There are four nodal regions provided by the standing wave of γγγγ in one dimer. As the lightest charged particles, electrons decouple the E2 field in the nodal regions (Fig. 1a), otherwise a kinetic contribution increases the total energy by momentum transfer. Two spinless Mott electrons gather together in the nodal regions between primitive cells while the highly gapped electrons in the filled shells are concentrated near the nuclear centres. Similar to a hydrogen atom, the Nb atomic structure provides the low-lying NKM groundstate of 5p dimers at ν < 1, where the 4d mixing vanishes in the large $N_I$ limit. The atomic and nuclear structures of Rh are more complicated than those of $^{93m}$Nb. The $^{103m}$Rh lotton may mix the p-wave and d-wave NKM orbits, caused by a level crossing [22]. The d-wave $^{103m}$Rh lotton has a heavy hole, which gives rise to a different $N_I$ according to (1), i.e., a different size of $^{103m}$Rh lotton.

Before the electrons complete the virtual photoelectric ejection by γγγγ, they are retrieved by the fast $\omega_R$ without any transfer of momentum. Using the language of quantum optics [26], the strong-coupling γ dresses the electrons rather than kicking them out, except for an incoherent internal conversion [63] without an intermediate γ together with some losses at the impurities and boundary [12]. We obtain $\hbar\omega_R = 84$ eV from $\hbar\omega_\gamma = 30.8$ keV. If we take $\omega_R$ to be the cyclotron frequency $\omega_{Nb} = eB_L/M_{Nb}$ of LLL Nb$^+$ ion, the instanton flux density $B_{IF}$ is $\sim 1.23 \times 10^{11}$ T. However, the pseudospin $r_+$ of $l_z = \pm 1/2$ yields the spontaneous instanton $B_+ \sim 3.5 \times 10^4$ T in the absence of LLs. The strong $B_+$ deforms the spherical atoms to provide a superradiance of K x-rays, which is caused by the indistinguishable internal conversions inside the solitons.

*1.2 The anisotropic pseudogap of the Fermi moat and the Higgs fields*

Underdoped AF ladders open an anisotropic pseudogap and reconstruct small pockets of the Fermi moat in the **k**-space. This is similar to the pseudogap at anti-nodal points in a underdoped cuprate of a high-$T_c$ superconductor, which reconstructs a d-symmetry Fermi surface with four small pockets. Chowdhury and Sachdev introduced a slow spacetime-varying Higgs field rooted in the collectivity of a spin liquid, which measures the local AF correlations in a rotating reference frame [65]. The real Higgs field revises the emergent gauges from case to case, which yields a more general meaning than just an amplitude mode of broken global gauge invariance [66].

We herein adopt the key concept that the real Higgs fields refer to nothing but the emergent flavours and their masses. A number of intertwined complications among three principal Higgs fields in a large $T$ must be clarified. The elementary NKM Higgs field $\langle\mathcal{H}_{NKM}\rangle \neq 0$ for the emergent $\omega_R$ mass gap pairs all kinds of fermions to the spin-0 dimers while confining two monopoles by a force $\propto n_L$. The major γ-energy characterized by $\omega_\gamma$ is nearly equal in a constant volume, as long as the changing $n_l \ll N_I$. AF ladders undergo BEC with $\langle\mathcal{H}_\gamma\rangle \neq 0$ at $\nu > \nu_{qc}$. The nuclear spin-density wave [15] Higgsed by $\mathcal{H}_{NKM}$ becomes high-order Higgsing fields to generate Majorana masses [22]. More specifically, the "gyration" of canted AF ladders provides the orbit-orbit coupling between the NKM and MRP/RRP orbits, which gives rise to the Majoranization by hyperfine splitting of the huge Dirac mass. The third $\langle\mathcal{H}_S\rangle \neq 0$ appears with the superconducting transition. The superconducting $T_c$ is independent of $\nu$ in several TME measurements. Therefore, the individual electron dynamics in the two sectors are nearly independent.

*1.3 The anomalous topological Kondo effect and the anapole condensate*

We observed an anomalous topological Kondo effect to show an anapole $T_{Ka} \sim 61$ K without applying $H_z$ at $\nu \sim \nu_{qc}$ (see Fig. B3 in Appendix B), i.e., a negative resistance and a longitudinal but no transverse TME, caused by a Lifshitz transition between flavours. We argue that the negative TME resistance to gaining energy along with a huge inductance is true, as detailed in Appendix B3. Therefore, there is a power source in the nontrivial vacuum to couple instantons.

The anapole condensate was found by lowering $H_z$ rapidly from 90 to 0 kOe at 100 K in the second test. The mutual thermal equilibrium between two heat reservoirs of electrons and coherent nuclei [67] was absent for at least a few hours. The nuclear NKM pseudospin temperature was very cold, where a large $n_L$ persisted. Had we reversed $H_z$ very rapidly, the Zeeman $T$ [67] would have become negative even at 300 K.

The RRP vortices detached themselves from the boojums, which were caught by the annular field from the drive current $I_x$. The unstable Dirac monopoles of nuclei at $T > T_{Ka}$ rejoined to form ring-type Dirac strings looping the drive current, which recovered the spontaneous $B_-$ but not the $B_+$. We now have a nonuniform distribution $B$ over the cell cross section, where the collective Mott electrons cannot see the nuclear Dirac rings carrying the fractional $2\pi/n_A$ flux. Almost all of the RRP $H_z^2$-magnets were maintained by the AF paired anapoles at the residual $H_z \sim 60$ Oe. The collective Peierls gap of nuclei at 90 kOe is greater than the spontaneous Kondo minigap of Mott electrons by five orders of magnitude. Hence, the relaxation of a large $n_L$ is exceedingly slow, and is characterized by a Korringa-like constant in metals [67]. Although the boojums needed to stretch RRP vortices vanish, the high-$n_L$ AF anapoles persist in the period to cool the sample from 100 to 72 K. The flavour transition also implies that the AF anapole groundstate has an energy slightly lower than that of AF ladders. We might have observed a spontaneous flavour transition in the presence of a Floquet drive by quenching $T < 77$ K at $\nu > 1$. However, the annular field cannot detach the AF ladders from the boojums at 4.2 K and $\nu < 1$ (see Fig. B1 in Appendix B), particularly when the AF ladders are parallel to the drive current (see Fig. B2 in Appendix B).

The continuous inductance as a linear function of $T$ and the negative quantum anomalous $T$-plateaux together validate the anapole condensate. The independent thermal equilibrium is attributed solely to the fast electron dynamics, which decouple the coherent nuclei. Three periodic drives of $\omega_\gamma$, $f_d$, and $m_a$ provide the complicated resistive structure, where $m_a$ is attributed to a very light spin-0 axion field [68]. New edge states of AF anapole dimers emerge from a commensurate condition, which will be discussed later on. Two negative plateaux regarding $T$, $f_d$, $\nu$ and *etc.* spanned by the $8\nu R_K/N_I$ reveal the octonion nature of the anapoles, which conserve the gofour flavours as an incompressible liquid while gaining energy from the axion field. Had we measured the transverse response at that time, there would have been no transverse TME at all, as demonstrated by the other observations undertaken at $\nu > 1$. More precisely, we would have observed a resistance between $\nu_5$ and $\nu_6$ (see Fig. 1d), which were caused only by a nonzero $I_x$ projection. This transverse resistance would have depended only on $T$ rather than $f_d$, $I_x$, $H_z$, and the rotational angle $\theta_R$, as illustrated in Fig. B2 of Appendix B.

Although the whole TME along with the cooling $T$ is detailed in Appendix B3, we view the subtle flavour physics from a different angle using a reversed time sequence, i.e., raising $T$ from 4.2 to 100 K in the thermal equilibrium of electrons. BPS $I_x$ flows uniformly in the sample at 4.2 K. AF paired anapoles restore the $\mathcal{T}$, PH, and chiral symmetries like their AF-ladder cousins at zero $H$. The anapole Kondo minigap closes to free Mott electrons from $B_L$, when $T > T_{Ka}$. However, a dynamical connection between two heat reservoirs remains as long as $T < T_{Ka} + 8hN_I f_d/k_B \sim$ 105 K, where $k_B$ is the Boltzmann constant. $I_x$ reopens sequentially four active gaps upon the thermally smeared minigaps to screen dynamically the EM polarizations of the nuclei. The cosmic spin-0 axion field [68] creates a topological line defect that penetrates the anapole dimer chain, inside which the whole $I_x$ contracts to maximize the system entropy. The Floquet drive breaks the $\mathcal{T}$-symmetry spontaneously to provide the continuous nuclear polarization of the anapoles by raising $T > T_{on2} \sim 72$ K, which contains a twist memory being similar to the case of Fig. B1 in Appendix B1. However, a distinct thermal potential $> T_{on2}$ is now required to release the screening of Mott electrons from the dynamical Floquet-Kondo singlets for the $\mathbb{T}^4 \times S^4$ solitons rather than releasing the screening of Cooper pairs for $S^4$ solitons at $T > T_{on1}$.

Three 7-tuplets emerge from the extremely sharp Mollow triplet in the electronic spectrum at $T_{on2}$, as provided by the fast Rabi oscillation and the slow $I_x$ modulation. Three central Mollow peaks merge their small neighbouring wings one by one, when raising $T$ sequentially by more than three onset temperatures to release the Floquet-Kondo screenings, i.e., $T_{on2}$ of the first type and the following two of the second type. The restored symmetries are broken in the nuclear but not in the electronic heat reservoirs. At the moment of breaking the symmetries at $T_{on2}$, three main peaks have already swallowed their nearest neighbours in three 9-tuplets, leaving the above-mentioned three 7-tuplets. The annular field of $I_x$ drives the breathing mode, where the annular ferromagnetic mixing flavour of parallel gofour anapoles, i.e., $\mathbb{T}^4 \times S^4$, in the AF dimer chain gives rise to the $T$-dependent inductance. Instead of driving $B$ to exhibit a gauge structure like the quantum Hall effects, we drive $T$ to show the resistive MZM $T$-plateaux. The axion wind supplies energy for the negative resistance. The increasing $T$ pumps the annular nuclear polarization $\propto (T - T_{on2})$ by reducing the Floquet-Kondo screening, where $T_{on2} = T_{Ka} + 2hN_I f_d/k_B$. The standard Kondo screening is nonlinear in $T$. Hence, the linear thermalization between 72 and 100 K is caused by the dressing gaps $2mhN_I f_d \ll k_B T_{Ka} + 2(m-1)hN_I f_d$, where $2 \le m \le 4$ denotes the magnetic quanta of Mott electrons.

Two kinds of $T$-plateaux, i.e., one smeared plateau at 90.5 K and two plateaux spanned by $8\nu R_K/N_I$ imply two kinds of small gaps. The first is the axionic gap $\Delta_a = 2hm_a$ and the second is the AF gap $\Delta_{AF}$, both of which are attributed to the pairing of clockwise and counter-clockwise anapoles. The AF gap is attributed to the spontaneous and dynamical EM polarizations of AF anapoles. The axionic field causes pairing of the AF anapoles by means of the chiral magnetic currents rather than by dressing the s-wave Cooper pairs [68]. We have observed a $\Delta_a$ fine structure at the exact commensuration of three periodic drives to estimate the axion mass, as discussed later on.

The anapole Kondo minigap $N_I (\mu_B B_L)^2/E_F = k_B T_{Ka} \approx 5.23$ meV gives a preliminary estimate of the lotton field $B_L \sim 17$ mT, where $\mu_B$ is the Bohr magneton and $E_F = 5.32$ eV is the Fermi energy. The normal resistance recovers by a lowering of $T \ll T_{Ka}$ at zero $H$ (see Fig. B3 in Appendix B). The free tunnel to cross the lotton wall is supported by a normal conductivity independent of $\nu$ at $T \gg T_{Kl}$. However, the situation is quite different in the presence of Cooper pairs.

A quantum-critical density of $\nu_{qc1} = 1$ is $1.935 \times 10^{12}$ cm$^{-3}$, which yields another $\nu_{qc} \sim 0.88$ of $1.70 \times 10^{12}$ cm$^{-3}$ to reflect the observed BEC near $10^{12}$ cm$^{-3}$ in the previous report [12]. Note that x-rays prefer emission along the long axis [14]. We previously estimated the $^{93m}$Nb density using x-ray counts along the short axis, which gave a low value.

The BEC critical temperature $T_{LC}(\nu)$ depends on $\nu$ and even on sample shape, but $T_{LC}(\nu) \to \hbar\omega_R/k_B$ together with a $T_{Kl}(\nu) \gg 61$ K is located above $\nu = 1$ while $T_{LC}(\nu) \to 0$ when $\nu \to \nu_{qc}$. Hence, the BEC transition is almost independent of $T$ but depends on $\nu$. Rather confusingly, it looks like a zero-$T$ quantum phase transition [47], as pointed out in a previous report [12]. It is not clear at present how $T_{LC}(\nu)$ develops except that one $T_{Kl}(\nu) \sim 225$ K was observed for $\nu > 1$, as detailed in *SI*.

The $^{103m}$Rh lottons have standalone anapoles without applying a Floquet drive current at $T < 77$ K and $\nu < 0.3\nu_{qc}$, which was not reported previously [10]. The $^{103m}$Rh lotton has a level crossing of NKM orbits at $\nu < 1$ whereas the $^{93m}$Nb lotton does not. Two species of lottons have quite different $B_L$ behaviours at $\nu < 1$, as discussed later on. The quenching of liquid-nitrogen (LN$_2$) creates spontaneously a peak-energy shift of the K x-ray spectra. We observed an abrupt reduction in $^{103m}$Rh x-rays to support the standalone anapoles, which are detached from the sample surface (see Fig. A3 in Appendix A). The x-ray intensity jumped back to the normal decay trend immediately, when the LN$_2$ cooling stopped. This abrupt increase in x-ray intensity reveals the restored boojums. In the meantime, the spectral shift recovered spontaneously but with an $\hbar\omega_R$ opened to a level slightly higher than 100 eV for the whole of the decay time down to $\nu \ll 0.3\nu_{qc}$. This spectral $T$-response reveals that the anapole gap is slightly greater than 6 meV.

*1.4 The decomposable spacetime*

The exchange of lotton chains leaves a 2D trajectory, which cannot shrink to a point. We thus decompose the (3+1)D into (2+1+1)D spacetime, which imitates string theory [24] at the low dimensionality, and provides rich nesting properties and an explanation for the anyon soliton in 4D. The lotton chain acts like a classical xy rotor with a specific angle $\varphi_L$ in the large $N_I$ limit, which advances as a Peierls phase in the 2D magnetic lattice. The classical 2D xy rotors equate to a 1D quantum rotor model to provide $\nu_{qc}$ [47].

We may stretch the lotton to a single string, which forms a nontrivial Su-Schrieffer-Heeger chain [9] of instanton dimers. The ion flux has $B_{IF} \sim 1.23 \times 10^{11}$ T in the nuclear string core. However, we cut the long string into $n_A \sim 3.5 \times 10^6$ pieces and fuse them to a shorter bundle of strings according to the $B_L \sim 17$ mT. The instanton $B_+$ pinned at the femtometer coil is now $\sim 3.5 \times 10^4$ T. The nonuniform field diverges very fast from $B_+$ to $> B_-$ in primitive cells, where the strong $B_+$ expands to the weak lotton $B_L$ by six orders of magnitude.

The nontriviality of a bundle of shortened Su-Schrieffer-Heeger chains becomes apparent in the $\mathcal{T}$-breaking Kondo triplet, which gives the chiral anomaly [1, 2, 24]

$$\sigma_{xy} = N_I/C_2 R_K, \qquad (2)$$

where $N_I$ comes from $N_I \alpha$. The anomalous Hall current $I_x$ on the x-y domain wall drives the lotton chains orientated along the y-axis (see Fig. 1d). $C_2 = 2$ or 4 is the semion number of lattice sites per magnetic unit cell, which reveals the 2- or 4-fold degeneracies regarding the generations.

The $2\pi$ NKM flux gives the lotton a cross section $A_L$ by $B_L = \Phi_0/A_L$ and a length $L_L$, where the $2\pi$ flux quantum is $\Phi_0 = 4.136 \times 10^{-15}$ Wb and $N_I$ atoms occupy a fixed volume $A_L L_L \sim 0.517$ μm³. Observation of the Arago spot indicates $\sqrt{A_L} \sim 0.3$ μm in the case of $v \gg 1$ [15], whereas we obtain a lotton diameter $\sim 0.5$ μm, $A_L \sim 0.24$ μm², and $L_L \sim 2.1$ μm at $v_{qc}$. The lotton aspect ratio is $\sim 4$. $A_L$ contains $n_A$ lattice sites and $L_L$ contains $n_L$ layers (Fig. 1b), which are tunable by $H_z$ but conserve $N_I = n_A \times n_L$. The length $L_L \propto H_z$ gives a relative permeability $\mu_r \sim 2.0 \times 10^6$ for the instanton $B_+ = \mu_r \mu_0 H_z$, which is greater than $M_{Nb}/M_e$, because the area of the primitive cell is greater than the area of electrons looping $B_-$. $N_I$ coherent ion orbits have a collective cyclotron gap $\hbar N_I \omega_{Nb} \sim 19.6$ eV/T. The spontaneous nuclear magnetic order is protected by the Peierls gap $\Delta_P \sim 0.3$ eV.

*1.5 Interface energy and morphology between two topologically distinct sectors*

When the interfacial energy between Cooper pairs and lottons is negative, AF ladders are dispersed uniformly in the Cooper-pair moat. In contrast, the positive interfacial energy groups the lottons together (Fig. 1e). The magnetic properties of type-I and type-II superconducting slabs have been well studied [48], but these become quite complicated in the presence of $\mathcal{T}$-dependent interface energy. The lotton Meissner effect of broken U(1)×U(1) symmetry becomes obvious in the absence of superconductivity at $T > T_c$. Many horizontal AF ladders remain untouched while $H_z$ only rotates a part of them to the edge LLL vortices. A lotton London length $\lambda_L$ screens $H_z$ by $\langle \mathcal{H}_{NKM} \rangle$, which is nearly independent of $T$ near $T_c$. Hence, $\lambda_L(T)$ equals $\lambda_{Nb}(0) = 39$ nm near $T_c$, where $\lambda_{Nb}(0)$ is the Nb London length at zero $T$. The ratio $\kappa = \lambda(T)/\xi(T)$ between a composite $\lambda(T)$ and a coherent length $\xi(T)$ depends significantly on $T$. The type-II superconductor becomes an effective type-1.5 superconductor without any supercooled normal state [48], where the interface energy vanishes at the BPS $T \sim 4.6$ K, as approximated by $\lambda(T) \sim (\lambda_{Nb}^{-1} + \lambda_L^{-1})^{-1}$.

The topological defects under the central Floquet drive current may become quite complicated, and far beyond our present understanding. For example, they depend on three parameters and the twist memory, their operational path, and even their tuning speed, i.e., two critical values of $T$ at 9.3 K and 4.6 K, several critical values of $H$, and the Floquet frequency $f_d$. Only one domain wall in Fig. 1d is frozen in the twist memory by lowering $T < T_{on1}$ ~ 8.1 K at the earth's field. Raising $T > T_{on1}$ using the same $f_d$ reopens the single domain wall at the earth's field. In contrast, a complicated domain-wall structure may emerge uniformly either at 4.2 K or at $T > T_c$, as shown later on. It is currently not well understood how the nontrivial domain-wall structure in a superconductor enables a voltage reading to be observed between the pickups (see Fig. 1c), which are two via holes apart from those used to feed the current.

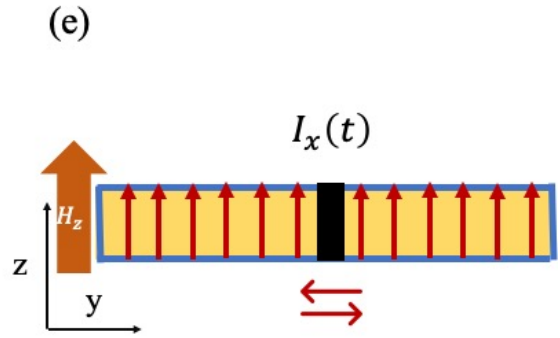

Figure 1, e) Morphology of PHP vortices at 9 K. The blue layer is the superconducting sector, which encloses the AF ladders (yellow) in the middle. The applied $H_z$ field excites a magnetic lattice of PHP vortices (red arrows), which are capped by the coaxial Abrikosov lattice (not shown). The applied ac current $I_x(t)$ using $f_d$ = 4 Hz creates the dynamic domain wall (black) at 9 K, where the bouncing vortices cut $I_x(t)$. The single domain wall and two superconducting layers disappear when lowering $T$ down to the BPS condition at 4. 2 K. Instead, a thin-film-like intermediate state freezes the PHP lattice as one of the nontrivial domain-wall structures.

*1.6 Longitudinal quantum anomalous plateaux at zero H*

We apply a current $I_x(t) = I_{ac}sin(2\pi f_d t)$ along the x-axis, similar to that shown in Fig. 1d apart from an incline angle $\phi_L$ ~ 50° between the MRP/RRP vortices and sea level. The Floquet drive breaks the flavour symmetry spontaneously and results in an opening of the domain wall at $T > T_{on1}$. The domain wall harbours the MZM, and allows a voltage between v$_3$ and v$_4$ inside the superconductor. The earth's field ~ 0.4 Oe lies at about 40° relative to sea level in Beijing. The AF ladders on the oval x-y plane are almost x-y-orientation degenerate. Hence, the AF ladders were roughly perpendicular to the earth's field (see Fig. B1 in Appendix B), where the y-axis was aligned to gravity and the x-axis to the NS direction.

Two longitudinal quantum anomalous $T$-plateaux of $2R_K/N_I$ and $4R_K/N_I$ of (2) as well as a spontaneous inductance at the particular $T$ intervals (see Fig. B1 in Appendix B) appeared only at 6 Hz among several values of $f_d$ ranging from 1 to 40 Hz. The $T$-plateaux reveal two incompressible quantum liquids of two distinct flavours in the stroboscopic sense [39] while the induction reveals the separation of AF ladders to provide the Weyl nodes. The LLL vortices acquired unidirectional phases $\pm\Delta\varphi_{LLL}$ in one period of a Floquet cycle [39]. After $4 \times 92n_H$ cycles, the vortices completed the full windings.

One MRP plateau appears below $T_c$ along with one domain wall sandwiched by two superconductors, and one RRP plateau appears above $T_c$ in an octonion intermediate state between the normal-conducting and superconducting phases, which probably constitutes a nontrivial domain-wall structure, being a superlattice of multiple domain walls and multiple superconducting sheets. We focus firstly on the MRP plateau. $I_x(t)$ charges the boojums of MRP $\pi$ fluxes to acquire the $V_x = 2cos(\phi_L - \phi_{IV})I_x R_K/N_I$ plateau $\approx 1.5$ μΩ according to (2). We obtained a value larger than the $cos\phi_L$ projection, as caused by the angle $\phi_{IV}$ ~20° between the v$_1$-v$_2$ and v$_3$-v$_4$ lines (see Fig. 1c).

The spontaneous opening of the domain wall depends on $T$ and $f_d$. The subtle relationship between $T$ and $f_d$ provides one of the key pieces of evidence for $N_I\alpha$. The energy of $N_I h f_d$ ~ 0.712 meV approaches the superconducting gap $\Delta_S(T_{on1}) \sim \Delta_S(0)(1 - T_{on1}^2/T_c^2)$ ~ 0.7 meV, where $\Delta_S(0)$ equals 3.05 meV and $T_{on1}$ is 8.1 K. It splits spontaneously two chiral cat states at $T$ > 8.1 K leading to the plateau related to $T$ and $f_d$. We did not scan $f_d$ in detail, but we note that the $T$ variation changes the matching point of $hN_I f_d = \Delta_S(T_{on1})$. A $\Delta T$ shifted the onset temperature $T_{on1}$ from 8.1 K to 8.6 K between two distinct measurements, where the cooling process contained a one-Hz periodic thermal gradient of $\tilde{T}(z,t)$ >> ± 0.1 K induced by the heat pump, while the heating process was free from $\tilde{T}(z,t)$ (see Fig. B1 in Appendix B). There would be no $\Delta T$ for a linear gap function of $T$. This $\Delta T$ demonstrates the change in matching condition $hN_I f_d = \overline{\Delta_S[T + \tilde{T}(z,t)]}$, which is caused by the increasing gap of the quadratic function in the presence of $\tilde{T}(z,t)$.

The sample entered the intermediate state provided by the periodic $\tilde{T}(z,t)$ kicks between 9.3 and 10 K. The pristine sample showed a wiggling roll-down without any TME passing $T_c$ (not shown), whereas the activated sample showed a plateau of $4cos(\phi_L - \phi_{IV})R_K/N_I \approx 3.1$ μΩ and an induction at $T_c$, where $I_x(t)$ charges the boojums of RRP $2\pi$ fluxes. Crossing $T_c$ improves the Q factor of RRP vortices by enlarging $\tau_{R1}$ to cause induction briefly during the cooling process, when the conduction loss vanishes.

In reality, 6 Hz is slightly higher than the exact commensurate frequency $4 \times 92 f_{LLL}$, which gives significant induction rather than being a purely resistive plateau. The relaxation time of a superconductor depends on $T$; however, the inductive contribution remained constant over the $T$-plateau. We thus know that $\tau_{R1}$ is the character in the nontrivial sector, which can be affected by electrons in the Fermi moat but not by the torsion-free Cooper pairs.

The periodic protocols of the instanton QTC and the Floquet drive share a common phase $\Delta\varphi_{LLL} = \omega_R/\omega_\gamma = n_H f_{LLL}/f_d$ per cycle, where $n_H$ =1 and $2\pi f_{LLL} \cong eB_L/N_I M_e$. We assume hereafter that the MRP $f_{LLL} \cong 15.625$ mHz and $4 \times 92 f_{LLL} \cong 5.750$ Hz using a better estimation, as explained later on. This assumption yields the estimates of $B_L = 16.01$ mT and $T_{Kl} \sim 56$ K in AF ladders rather than 17 mT and $T_{Ka} \sim 61$ K in AF anapoles. We thus adopt the $B_L$ shift from 16.01 to 17 mT to be important evidence of anapole formation. The estimates obtained using $B_L = 17$ mT in the preceding section, e.g., $A_L$, $L_L$, and $\mu_r$, must be modified accordingly for AF ladders.

The near-commensurate constrain of $|m_H f_d - 4n_H \times 92 f_{LLL}| < 1/\tau_{R1}$ with $\tau_{R1} \sim 2$ seconds rejects $f_d = 22$ Hz and $f_d = 3.3$ Hz to open the domain wall (see Fig. B1 in Appendix B). Instead, the twist memory is frozen in the nontrivial sector, when lowering $T < T_c$ using the near-commensurate $f_d = 6$ Hz. Note that the AF ladders are non-associative before lowering $T < T_c$. The twist memory wakes up, when increasing $T > T_{on1}$ regarding the Cooper pairs in the trivial sector of the type-1.5 superconductor.

We assume that RPP $f_{LLL}$ is 7.813 mHz. The induction vanishes where a smaller RRP Q quality factor was caused by the reduction in $\tau_{R1}$ in the presence of an Eddy loss, as shown by another case later on. The RRP plateau of $4cos(\phi_L - \phi_{IV})R_K/N_I$ of $n_H = 2$ free from induction showed up in the presence of $\tilde{T}(z,t)$ and the nontrivial domain-wall structure, whereas the MRP plateau of $2sin(\phi_L - \phi_{IV})R_K/N_I$ of $n_H = 1$ did not survive $\tilde{T}(z,t)$.

*1.7 Longitudinal quantum plateaux at nonzero $H_z$*

The Abrikosov vortices were spread all over the sample near $H_{c2}$ (Fig. 1e), and we must consider the housing problem between the PHP/MRP and the Abrikosov vortices. Two kinds of vortices in the Weyl superconductor have been noted [54]. The first kind is a coaxial case near $T_c$ in the weak $H_z$ limit. The superconducting layer at the sample boundary encloses AF ladders in the type-I superconductor (see Fig. 1e). The π fluxes penetrate the PHP vortices as sandwiched by two top-down caps of Abrikosov vortices. The second kind is either a non-coaxial case or a different coaxial configuration altogether. The Abrikosov and MRP π vortices coexist either at distinct positions or at the same coaxial positions, but this leaves room for untouched horizontal AF ladders to penetrate the bulk.

Following each new setting of $H_z$, we had to wait until the reading became stable, as caused by a remarkable relaxation time of the slow thermal mixing between the electronic and nuclear interaction reservoirs [67].

We observed a pronounced TME hysteresis caused by the Lifshitz transition to distinguish the $H_z$-up and $H_z$-down phases below the critical field $H_{c1} \sim 4$ kOe of the type-1.5 superconductor at 4.2 K, i.e., either the zero-TME superconductor in the $H_z$-up phase or the $H_z^2$-supermagnets in the $H_z$-down phase (see Fig. 2). The drive current should flow uniformly around the BPS lottons to form a nontrivial domain-wall structure. The Lifshitz transition between flavours preserves the octonion twist when lowering $H_z < H_c$, and these provide the $H_z^2$-supermagnets of torsion-free MRP vortices pinned by boojums. The quadratic power is attributed to the edge states of $n_L \propto H_z$ and the MRP number $\propto H_z$. The detachment and the reattachment of MRP vortices give rise to a hysteresis by looping between 2 and 4 kOe. The lotton Meissner effect tripled the pristine $H_{c1}$ ~ 1.4 kOe [69, 70] to achieve 4 kOe (see Table 1). The $H_{c3}$ ~ 12 kOe recovered to a normal value as in a pristine sample [69, 70] to reveal that the boojum sheath gradually detached itself from the boundary at $H_z > H_{c1}$.

The interface energy becomes > 0 because $\xi(T)$ diverges near $T_c$. The Lifshitz hysteresis disappeared at 9 K, where $H_{c1}$ became normal and instead $H_{c3}$ was abnormal (see Table 1). The Dirac nodes along the x-axis were preserved in the weak $H_z$ limit, as revealed by the zero induction below 500 Oe. The superconducting sector enclosed the nontrivial sector of AF ladders (see Fig. 1e). More precisely, the AF ladders and the vertical boundary sandwiched the surface Cooper-pair sheath from two sides along the y-axis, the particular boundary conditions of which survived a high $H_z$, as revealed by the tripled $H_{c3}$ at 9 K. The drive current might open only one domain wall using $f_d$ = 5.8 Hz instead of 4 Hz in the weak $H_z$ limit. However, we must bear in mind that the domain-wall structure depends on the operational path of parameters $T$, $H_z$, $f_d$, and the twist memory. The Abrikosov π-flux lattice eventually filled the sample near $H_{c2}$. The coaxial PHP vortices were provided by canted AF ladders lying in the x-y plane, which preserved an exact PH symmetry with degenerate $l_z = \pm 1/2$ orbits in the presence of Dirac nodes in the **k**-space [71].

Two distinct longitudinal $H_z$-plateaux appeared with regard to the distinct flavours, i.e., a resistive $H_z^0$-plateau of $4\nu R_K/N_I$ and an inductive plateau $\propto H_z^2$ (see Figures 2-4). The incompressible liquid conserves the PHP flavour near $H_{c2}$ to provide an $H_z^0$-plateau, which has a 4-fold degeneracy in the magnetic unit cell known as the Thouless-Kohmoto-Nightingale-Nijs invariant [72] rather than the $C_2$ in (2). The moving PHP/Abrikosov vortices cut $I_x$ to provide the quantum resistance $4\nu R_K/N_I$, which was reduced by a factor of ν due to the superlattice periods, as highlighted by the Haperlin formula [73]. In contrast, two distinct flavours are conserved at a specific $H_z$, which provides the $H_z^2$-supermagnet regarding various parameters.

All horizontal AF ladders were rotated by $H_z$, when the negative magnetoresistance vanished between 16 and 18 kOe at 10 K. The same negative trend also appeared but at a higher level when the lottons passed $H_{c3}$ at 4.2 K. Generally, the resistance at 4.2 K was lower than that at 10 K except for the Kondo effect. Horizontal AF ladders and vertical RRP vortices coexisted below 17 kOe. The $\mathcal{T}$-breaking magnetoconductance increased linearly [32], which led to the common Kondo effect between 0 and 16 kOe at 8, 9, and 10 K.

We observed a longitudinal impedance ~ 4 mΩ at 90 kOe and 10 K, which enhanced the residual resistance 4.5 μΩ at zero $H$ by a factor of about 1000 to demonstrate a metal-insulator transition caused by the flux lattice. The quantum plateau of 3.16 μΩ gives an estimate of ν ~ 0.88. The TME disappeared one month later. In the meantime, the quarter γ disappeared too. We thus obtained $v_{qc}$ ~ 0.88.

*1.8 The lotton axion electrodynamics*

We now highlight the $H_z^2$-induction, in particular at $H_z > 17$ kOe (see Fig. 4), which corresponds to the TME of an A-type Floquet Chern insulator [74] embedded in the Weyl metal. The Kondo minigap levels at high $H_z$ are well separated without overlapping for $T_{Kl} \gg 300$ K. The current density $J_x(t)$ flowed uniformly through the sample with $f_d \ll 4 \times 92 f_{LLL}$ ranging from 4 to 10 Hz, which was lower than a resonant STB frequency near 20 Hz, as explained in the discussion. The $H_z^4$-terms are mainly attributed to a nontrivial Eddy screening in the Fermi moat and the spectral flow of Mott electrons (see Table 2), as revealed by a significant reduction at 100 K. We leave these $H_z^4$ terms regarding the $S^8$ STB dynamics of octonions to a future investigation.

The chiral edge current $J_x(t)$ in the Fermi moat induces an electric field $\boldsymbol{E}$ in the RPP cat states. Here, we focus on the axion term $E_z \propto J_x$ and the $H_z^2$-inductive part of $E_x \propto 2\pi f_d J_{AH}$, while leaving for instance the $H_z^2$-resistive part in $E_x$ for a later discussion, acknowledging that it nevertheless shows a strange STB $f_d$ response. The presence of $E_z$ introduces the anomalous Hall current density $J_{AH} \propto E_z H_z^2$ in the nontrivial sector to generate $E_x$ induction. Two cascade chiral anomalies work like a $H_z$-dependent transformer converting one anomalous Hall current $I_x(t)$ in the trivial sector to another anomalous Hall current $I_{AH}(t)$ in the nontrivial sector. The TME $\propto H_z^2$ becomes manifest using a 2D×1D product, where RRP vortices have a density $n_{xy} = 4\mu_0 H_z/\Phi_0$ on the x-y plane and a zero-mode density $n_L = 4N_I\mu_0\mu_r H_z/B_{IF}$ in the k$_z$ space. We obtain

$$J_x(t) = N_I E_z(t)/C_2 R_K, \tag{3}$$

and

$$J_{AH}(t) = n_L n_{xy} N_I E_z(t)/C_3 R_K \tag{4}$$

regarding the axion electrodynamics in the two sectors, where $C_2$ and $C_3$ are equal to 4 for RRP. The chiral anomaly in (3) is revised from (2). The non-abelian pentagon diagram in (4) contains the $N_I^2$ factor [31], one $N_I$ of which is hidden in $n_L$.

We obtained 5.5 fH/Oe² from the $H_z^2$-supermagnet at $H_z < 4$ kOe, $f_d = 4$ Hz, and 4.2 K, which occurs in the *H*-down branch of the hysteresis between 0 and 5 kOe (see Fig. 2). The $H_z^2$-inductance develops slowly from 6 to 16 fH/Oe² by increasing $H_z$ from 4 to 16 kOe at 4.2 K, which becomes stable for $H_z > 16$ kOe (see Fig. 4 and Table 2). Faraday induction gives $V_x = -\int_{v_3}^{v_4} \partial_t A dl$, and the distance $d_{34}$ between the two via holes of v$_3$ and v$_4$ is ~ 8 mm (see Fig. 1c), with a vector potential $\nabla^2 A = -\mu_0 J_{AH}$. The stable $H_z^2$-inductance of 16 fH/Oe² ~ $\mu_0 d_{34} N_I/H_{zs}^2$ becomes apparent at $H_{zs} = B_{IF}/4\mu_r\mu_0$, where the magnetic unit cell consists of 4 fully stretched semionic vortices and $\mu_r \sim 2.13 \times 10^6$ is revised according to $B_L = 16.01$ mT. The RRP gyration adjusts the flux density from $B_{IF}$ to $B_{IF}/4$ to yield $n_L = N_I$ for the fully stretched lotton strings. When we lowered $H_z$ using coarse steps from 20 kOe down to 0 kOe, the TME did not survive in the $H$-down branch at $H_z < 4$ kOe, i.e., the TME depended on the speed to tune $H_z$. Furthermore, the slight magnetization approaching 4 kOe in the $H$-up branch (see Fig. 2) also supports the lotton Meissner effect at the boojums to reduce the $H_z^2$-supermagnet.

*1.9 Lotton applications*

The quantum computation requires a basic qubit element, i.e., the superposition of 0 and 1 states. The non-abelian MZMs of the same zero energy provide the topologically protected qubits [43]. MZMs are usually stable at very low temperature, for instance mK [75]. The non-commutative lotton quaternions provide a straightforward way to process the soliton-protected quantum information. The neat BPS MRP response @ 4.2 K implies robust quaternions with extremely long quantum coherence. Furthermore, room-temperature MZMs of gofour anapoles are possible by means of tunning some parameters, for instance advancing $f_d$ from 4 to 40 Hz and pre-polarizing $H_z$ from 90 to 250 kOe as in Fig. B3 of Appendix B. Non-associative octonions above 300 K may offer a novel direction for the quantum computation.

We observed a laser-beam-like γ-ray near 7.5 keV at the earth's field, which is supported strongly by sub-Poissonian photon statistics [13]. This quarter γ ray was mistaken for an impurity Ni Kα x-ray in the first instance [13] but was shown to be one of the γγγγ later on, for two reasons [14], namely photon attenuation caused by the insertion of filters (for instance see Fig. A2 in Appendix A) and the absence of Kβ when the background noise of the oval single-crystal sample dropped spontaneously to a very low level one year later (in 2016). The coherent quarter γ and the commensurate relationship with $f_d/f_{LLL}$ provide many useful applications, e.g., a low-energy γ laser and a time standard [33].

The pseudoscalar axion $a_{DM}$ is a promising candidate for cosmic dark matter [68], with a mass $m_a \sim 10^{-4\pm1}$ eV [76], a local density $\rho_{DM} \sim 0.3$ GeVcm⁻³ [77], and an axion wind [78] related to a galactic virial velocity ~ $10^{-3}$ $c$. The ultrahigh $B_+$ favours the conversion of $a_{DM}$ to EM energy, where the Primakoff efficiency seems proportional to $B_+$ by convention [79]. In fact, the gofour anapoles provide a much higher efficiency $\propto N_I B_+$ than that of the ABRACADABRA toroid [78], as discussed below. The axion-γ coupling of $\propto N_I \alpha$ enhances the axion conversion into γγ [76] by $N_I^2$. It is highly possible to observe "light shining through walls" [80] of MeV γ [15].

The huge Kopnin mass amplifies the sensitivity to detect weak GWs by $N_I^2$. It is further enhanced in the $^{103m}$Rh case (see Fig. A4 in Appendix A), where the effective mass of nuclei is much greater than that of atoms. We will discuss the mixed chiral anomaly later on, where the spin-2 GW replaces two spin-1 legs in the box diagram. Hence, it is possible to detect the wave from one particular source like a GW telescope by tuning some parameters, e.g., the array-detector orientation and the matching condition.

The lotton physics may contribute a different perspective for understanding the strong *CP* problem of QCD [81], the enigma pseudogap of the high-$T_c$ superconductor, and the 5/2 quantum Hall plateau. The interplay among inhomogeneous sectors, indistinguishable quasi/particles, supersymmetric degrees of freedom [24], gauge fields, flavours, and many other well known phenomena [23] all give rise to plenty of topological objects, which in turn reveals the rich tapestry of topological quantum matter, some of which we highlight in this report.

2.  **TME experimental Results**

*2.1 Introduction and some GW signals*

The TME experiments were undertaken four times. The methods and the materials used for the experiments were detailed in a previous report [12]. We have now reassessed the results of the first and second tests, as well as some results from the fourth test. Two single crystals of Nb [a (110)-oriented oval plate, 1.2 mm × 12 mm × 13 mm] were used. The first sample was prepared by neutron irradiation in the reactor at Tsinghua University, Hsinchu, in June 2008; this lost its TME response one month later after the second test, and was therefore reactivated in November 2011. The second sample remained pristine until November 2011, when it was activated together with the first.

The first test was undertaken in January 2011 in Beijing and the second was undertaken in June 2011in Hsinchu, when the $^{93m}$Nb excitation density was about to pass $\nu_{qc}$ ~ 0.88. The third test was undertaken in November and December 2011 in Hsinchu, when $\nu \gg 1$. In the fourth test between November 2012 and January 2013 in Beijing, $\nu$ was estimated to be slightly > 1 because of the fast decay at $\nu \gg 1$. Three years later in 2016, we observed an abrupt change in x-ray spectrum particularly below 16 keV at room temperature, which might indicate a quantum-critical transition at $\nu$ = 1. We continued to monitor the decay until 2021. The characteristic quarter γ near 7.5 keV is still available. The decay is likely to be slower than the normal half-life at ν < 1.

We drilled six holes in the sample to apply standard four-point and Hall configurations, as shown in Fig. 1c. We applied indium solder to fix the Cu wires in the via holes in the first, second, and third tests. The wiring was improved using tight-fitting gold leads in the fourth test, in which we could rotate the sample under *H*. We measured only the longitudinal term in the first and second tests, but measured both the longitudinal and transverse terms in the third and fourth tests. The topology of the sample surface is a torus of genus 6. The sample had a somewhat asymmetrical shape (not shown in Fig. 1c) because a small slice was cut from the oval disc in order to measure the Meissner effect. The asymmetric shape is irrelevant to the longitudinal TME in this report, which exhibits a perfect $\pm H_z$-symmetry for $|H_z| > 4$ kOe at 4.2 K (not shown), as shown in the second test undertaken in Hsinchu.

In the first test undertaken in Beijing, the sample y-axis was along the direction of gravity. No external magnetic field was available, except that the earth's field in Beijing is oriented at 40° to sea level. We applied two 1:10 toroidal transformers to feed the ac current and to pick up the differential voltage that appeared at the two output leads. The frequency response of the toroidal transformer was flat, ranging from 2 to 600 Hz. The two lines of $v_1$-$v_2$ and $v_3$-$v_4$ crossed each other, as shown in Fig. 1c, in the first test.

The two lines of $v_1$-$v_2$ and $v_3$-$v_4$ were nearly parallel (not shown) in the second test. The sample x-y plane lay horizontally in the dewar, while the x-axis was along the N-S direction and the applied field $H_z$ was in the direction of gravity in Hsinchu. The AF ladders lay horizontally on the x-y plane along the y-axis, i.e., perpendicular to the residual $H_z \sim 60$ Oe and the earth's field $\sim 0.4$ Oe along the x-axis.

At the beginning of the second test, we scanned $T$ and $f_d$ to search for the $T$-plateau found five months previously. We chose $f_d = 10$ Hz and $I_x = 13$ mA to avoid the $T$-plateau (see Fig. B1 in Appendix B) while monitoring the resistance that was reducing with $T$. Surprisingly, a thin-film-like intermediate state [48] emerged in the mm-thick sample. We launched the PHP plateau of $4\nu R_K/N_I$ for a while below $T_c$ (see Fig. 1e). The TME vanished abruptly at 8.74 K. The true sample $T$ was slightly greater than the read values in the $T$-down phase, caused by delays in the heat transfer and the lock-in amplifier. Table 1 lists $H_{c1} \sim 60$ Oe at 9 K. The conventional $H_{c1}$ is 120 Oe at 9 K [69. 70]. We thus assume a residual $H_z \sim 60$ Oe, which will be verified by the PHP $f_{LLL}$ later on.

We then scanned $f_d$ from 2 to 30 Hz at 4.2 K. A magnet $\sim 36$ nH emerged from a nontrivial domain-wall structure under the BPS condition. More precisely, the PHP-magnets had an TME amplitude $\propto f_d$ and a TME phase between $\pi/3$ and $2\pi/3$. The phase increased with an arbitrary $f_d$ between 2 and 6 Hz, which gave an overshot $\sim 2\pi/3$ at 6 Hz. The swinging phase became stable at $\pi/2$ for $f_d \geq 10$ Hz, i.e., a PHP-supermagnet in the type-1.5 superconductor. There was a small inductance bump $> 36$ nH near 5 Hz. A 6-Hz third harmonic $\sim$ nV emerged significantly at 2 Hz, whereas the second and third harmonics of 3 Hz were insignificant.

The TME results showed two different values at the same condition using 10 Hz at 4.2 K within ten minutes. The former yielded a zero TME while the latter yielded a nonzero TME, which depended on the path and the rate of change of $f_d$ and $T$. We introduced this problem, wherein the TME may show path-dependent results for the same parameter settings. The PHP-magnet persist throughout the arbitrary $f_d$ scanning at 4.2 K.

In contrast, using $f_d$ = 3.3 Hz it was not possible to create the PHP-magnet in the first test (see Fig. B1 in Appendix B). The key difference between the two tests lies chiefly in the residual $H_z$ in the second test, which is two orders of magnitude greater than the earth's field. The phase transition closes the top-down Abrikosov lattices (see Fig. 1e) at $T$ < 8.7 K. The magnets in the absence of Abrikosov lattice at 4.2 K are thus attributed to PHP vortices fixed by the boojums along all three x-, y-, and z-directions. By estimation, $4 \times \nu \mu_0 d_{34}$ yields 36 nH for the PHP-supermagnet, where 4 is the groundstate degeneracy of the PHP flavour and $d_{34}$ ~ 8 mm is the distance between v₃ and v₄ in Fig. 1c. It took us nine years to understand that this peculiar non-associative physics is only caused by the small residual $H_z$ ~ 60 Oe. We note repeatedly multiple values for the same driving parameters throughout this report.

Octonions remember the twists at various critical points. For example, the TME at 4.2 K using 4 Hz was the PHP-magnet in the day-one study. It became zero TME in the day-three and day-four studies, as shown in Figures 2-5. The twist memories depend on the parameters entering the superconductive phase. Hence, the near-commensurate $f_d$ = 10 Hz with $n_H$ = 2 and $m_H$ = 1 memorizes the twist in the day-one study but the incommensurate $f_d$ = 4 Hz does not memorize the twist in the day-three and day-four studies. The exact $4 \times 92 f_{LLL}$ changes from 5750 ± 3 mHz in the first test to 4863.9 ± 0.2 mHz in the second test because of the residual $H_z$. We obtain $\tau_{R1}$ between 1.1 and 3.5 seconds from the second test and between 1.2 and 4 seconds from the first test, according to the constraint $|m_H f_d - 4 n_H \times 92 f_{LLL}| < 1/\tau_{R1}$. We thus obtain $\tau_{R1}$ ~ 2 seconds.

We scanned more values of $f_d$ and $T$ to investigate the induction bump. The exact commensuration was identified near 4.9 Hz, i.e., the phase moved from a capacitive TME to an inductive TME between 4.6 and 5 Hz by keeping $T$ near 8.3 K. Surprisingly, the TME did not hold on to a single value even when keeping $f_d$ and $T$ the same. To show this, the TME using a fixed $f_d$ = 4.89 Hz resulted in multiple values at 4.2 K, which are a time-varying resistance exhibiting a sawtooth modulation (see the left panel of Fig. B4 in Appendix B). The sawtooth evolution caught the passing GW at 0.53 mHz, showing constant peak-to-peak amplitudes ~ 0.8 μΩ. Hence, the resonant $m_H f_d \cong 4.8903$ Hz is nothing but the exact frequency $f_0 = 4 n_H \times 92 f_{LLL} + 2 n_G f_{LLL}$ of $m_H = n_H = n_G = 1$ to flip a seesaw quadrupole edge state in cooperation with a second GW Floquet drive, as discussed later on. We applied the $2 f_0$ and $f_0/2$ drive frequencies to double check this exact commensuration (see the right panel of Fig. B4 in Appendix B).

We scanned $f_d$ between 4.898 and 4.883 Hz, and thus picked up different GWs in a discrete manner (see the left panel of Fig. B4 in Appendix B). The GW frequencies did not mix, i.e., only one GW was dressed on the MRP edge state each time. The TME is like a capacitor, i.e., the amplitude becomes smaller for a higher frequency. For example, the peak-to-peak amplitude was 100 nV at 0.5 mHz but 60 nV at 6 mHz.

We studied the signal at 0.53 mHz using 4.889 Hz by changing $T$ from 4.2 to 10 K (see the right panel of Fig. B4 in Appendix B). The vanishing superconductivity abruptly shifted the signal by a residual resistance of 4.5 µΩ. The GW signal persisted even in the absence of superconductivity except that it propagated backwards in time, which was caused by the flavour change. These are the most important pieces of evidence for claiming the GW detection, for two reasons. Firstly, the EM waves cannot penetrate the superconductor. Secondly, the flavour transition changes the Majorana mass to couple GWs. A similar signal ~ 0.5 mHz emerged using 4.889 Hz, when we raised $H_z$ = 10 kOe at 4.2 K next morning in the day-two study. The lotton Meissner effect protected the same $f_0$ of PHP vortices at $H_z$ < 16 kOe.

We postponed our attempt to study the details of these peculiar signals; instead, we studied the pronounced TME by varying $H_z$. Unfortunately, the following month the sample was no longer active, and we must now wait for some decades in order to reach the unique condition of ν ~ $ν_{qc}$ after the reactivation.

We carried out calibrations to identify the system automatically by very slow heating from 4.2 K to 20 K with drive frequencies of 4 and 10 Hz every night, which provided the systematic biases for the measurements. We calibrated the measuring system carefully using these data. Assuming a zero resistance at 4.2 K, a pure residual resistance at 10 K, and a pure resistive transition in between, the systematic biases depending on the drive frequency in amplitude and phase are resolved from the complicated octonion memory. The residual resistance at 10 K was then updated from the previously reported 4.3 µΩ [12] to 4.5 µΩ.

As mentioned in the preceding paragraph, AF ladders remember the state for entering the superconducting phase. The following measurements in the day-three and day-four studies entered the superconducting phase using 4 Hz and 10 Hz, which provided stable capacitive biases of -10*i* and -25*i* nV in Figures 2-5, respectively.

*2.2 Longitudinal TME at low $H_z$*

Figure 2 shows the longitudinal TME with $f_d$ = 4 Hz at 10 and 4.2 K, which was driven by an ac current $I_x(t) = 13 sin(2\pi f_d t)$ mA through the x-axis under $H_z$. A significant $H_z^2$-inductance became apparent on increasing the field to 20 kOe at 10 K. The residual resistance of 4.5 µΩ decreased linearly for $H_z$ < 17 kOe, because the positive magnetoresistance only dominated in the high $H_z$ region. We observed a 4% Kondo effect between 0 and 17 kOe at 10 K. The same negative magnetoresistance appeared at $H_z$ > $H_{c3}$ at 8 and 9 K (see Fig. 3), whereas the Kondo effect showed up at a higher level for 13 kOe < $H_z$ < 17 kOe at 4.2 K. The negative magnetoresistance disappeared altogether at 100 K, particularly at $H_z$ < 2 kOe. Hence, $T_{Kl}$ ~ 56 K (see Fig. B3 in Appendix B) lies in the interval 100 K > *T* > 10 K at zero $H_z$.

The superconducting TME at 4.2 K demonstrated a pronounced hysteresis below 4 kOe (Fig. 2), i.e., insignificant TME in the *H*-up phase when increasing $H_z$ from 0 to 5 kOe, whereas a pure $H_z^2$-inductance decreased continuously in the *H*-down phase when decreasing $H_z$ from 5 to 0 kOe. This highly meaningful observation indicates the nontrivial domain wall, which gives a nonzero voltage in a superconductor. However, the hysteresis disappeared when we decreased $H_z$ from 20 instead of 5 kOe. The $H_z^2$-inductance ~ 4 kOe in the $H$-up phase is barely identifiable, while the resistive contributions in the up and down branches of the hysteresis were less than the resolving power of the measuring system ~ 2 nV for $H_z$ < 4 kOe. The $H_z^2$-supermagnet gives an estimate of 5.5 fH/Oe², which slowly develops to 16 fH/Oe² when $H_z$ > 16 kOe.

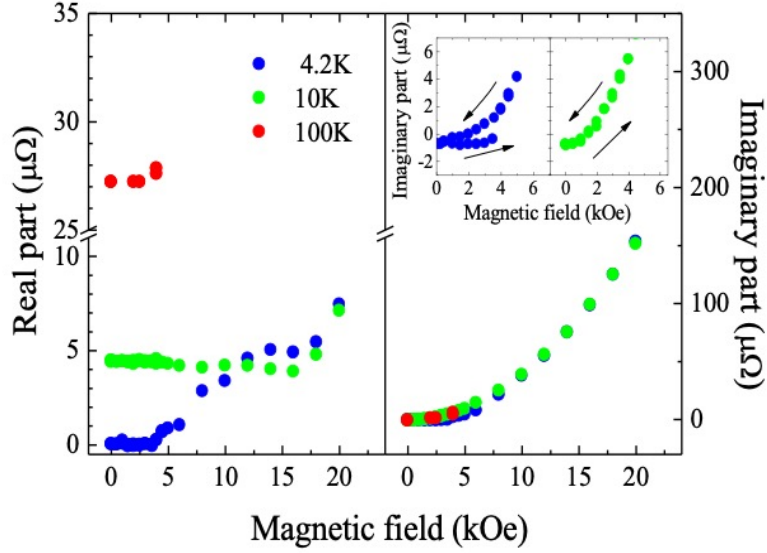

Figure 2, Resistive and inductive parts of the longitudinal TME at 10 and 4.2 K, which show two types of measurements. The first uses coarse steps between 0 and 90 kOe and the second employs fine steps between 0 and 5 kOe. The inset shows the hysteresis of the $H_z^2$-inductance between 5 and 0 kOe at 4.2 K whereas there is none at 10 K. The 10-K resistance decreases linearly with $H_z$, which returns to a positive trend above 16 kOe. A negative magnetoresistance appeared between 13 and 17 kOe at 4.2 K, which had a larger resistance than that at 10 K due to Kondo scattering.

The BPS $H_z^2$-supermagnets show hysteresis by looping $H_z$ between -4 and -2 kOe. For example, the induction at -3 kOe in the $|H_z|$-up phase was greater than that in the $|H_z|$-down phase. We had to wait a couple of minutes, because the reading oscillated unusually after a stepwise advance of 0.1 kOe. Hence, the inversed hysteresis at -3 kOe might be caused by the speed of tuning of $H_z$.

We adopt $H_{c1}$ = 4.0 kOe at 4.2 K, when the inductance abruptly emerges (see Fig. 2). The measured resistance was lower than the residual resistance of 4.5 μΩ at 4.2 K in the range of $H_z$ < 12 kOe. We thus adopt $H_{c3}$ ~ 12 kOe at 4.2 K. We compare the measured critical fields with the documented values [69, 70] in Table 1. The overly large critical fields of $H_{c1}$ at 4.2 K and $H_{c3}$ at 9 K (see Fig. 3) reveal lotton diamagnetism.

A pronounced plateau of 3.16 μΩ of $4\nu R_K/N_I$ between 0.1 and 1.0 kOe appeared at 9 K, as shown in the inset of Fig. 3.

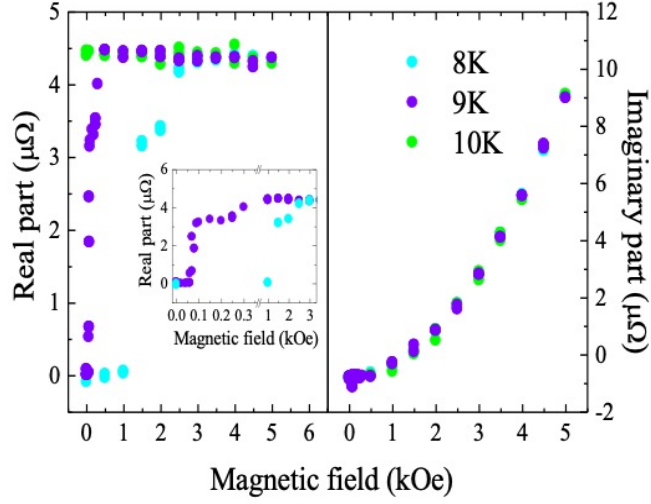

Figure 3, Resistive and inductive parts of the longitudinal TME at 8, 9 and 10 K. The inset shows a pronounced plateau of 3.16 μΩ at 9 K. The same $H_z^2$-inductance showed up for all $T$ settings, except < 0.5 kOe at 9 K.

| Temp. (K) | $1-(T/T_c)^2$ | Ref. $H_{c1}$ (kOe) | $H_{c1}$ (kOe) | Ref. $H_{c3}$ (kOe) | $H_{c3}$ (kOe) |
|---|---|---|---|---|---|
| 4.2 | 0.80 | 1.43 | 4.0 | 12 | 12 |
| 8 | 0.26 | 0.47 | 1.0 | 1.0 | 3 |
| 9 | 0.063 | 0.12 | 0.06 | 0.14 | 0.5 |

Table 1: Referenced and measured critical fields. *Ref.* $H_{c1}$ is taken from the reference [69] while *Ref.* $H_{c3}$ is taken from reference [70].

## 2.3 Longitudinal TME at high $H_z$

Figures 4 and 5 show two groups of 5 experimental results at high field, which are illustrated with a $H_z^2$ horizontal ordinate to identify two different $H_z^2$ and $H_z^4$ behaviours, i.e., the $H_z^2$ term is a straight line and the $H_z^4$ term becomes a parabolic curve. In Fig. 4 results were obtained with the 4-Hz drive at 4.2, 10, and 100K. For Fig. 5 the values were 4, 6, and 10 Hz at 4.2 K. We estimated the coefficients depending on $H_z^n$ with $n$ = 2, and 4, as summarized in Table 2. The zero-field resistances are fixed in the parameter fitting.

The Faraday inductions of $c_{i2}$ with $f_d$ = 4 Hz were almost the same for the three $T$ settings. This $T$-independent induction also supports the flavour transition from the RRP vortices to the mixing of AF and goufor anapoles, which have the $T$-dependent induction, (see Fig. B3 in Appendix B). The positive $c_{r4}$ and the capacitive $c_{i4}$ decreased markedly at 100 K, as caused mainly by the conduction in the Fermi moat and the spectral flow of Mott electrons. The zero-field resistance $c_{r0}$ increased by a factor of six, leading to a 14-fold reduction of $c_{r4}$ between 10 K and 100 K, which was larger than the Eddy loss $\propto 1/c_{r0}$. The ratio of $c_{i2}^2/c_{r0}$ gives a good estimate of Eddy loss, however, being significantly greater than $c_{r4}$ with a factor of ~ 3000 at 4 Hz @ 4.2 and 10 K. This provides evidence of the nondissipative channels of the chiral magnetic and anomalous Hall currents in the nontrivial sector. The Eddy loss in the magnetic lattice is reduced, just like a layered magnetic core in a commercial transformer. Neglecting the $c_{i2}^2/c_{r0}$ term, the $c_{r2}^2/c_{r0}$ term approaches $c_{r4}$ much more closely, but is still greater than $c_{r4}$ by a factor of ~ 3. The RRP vortices were adhesively pinned by the boojums to exhibit STB motion, such that the $c_{r4}$ dissipation to screen the spectral flow in the trivial sector was reduced. The $c_{r4}$ reduction between 10 to 100 K does not match the reduction between the $c_{r2}^2/c_{r0}$ terms. Some missing terms beyond the Eddy effect are required. The quadratic magnetoresistance $c_{r2}$ may in principle be attributed to the spectral flow, which is the same at 4.2 and 10 K but not 100 K, which implies that there must be other contributions, e.g., STB polarization.

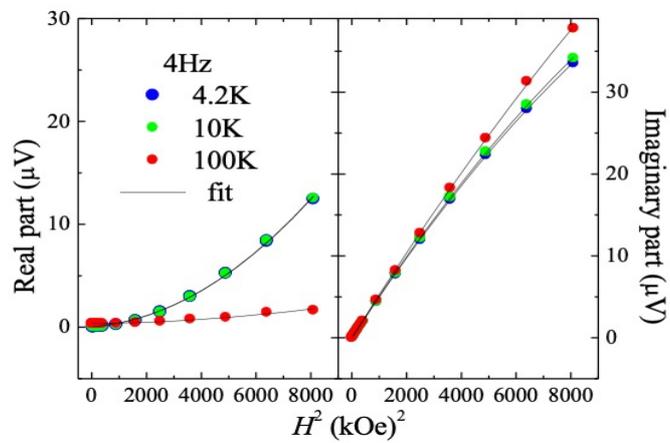

Figure 4, Longitudinal TME at various $T$, 4 Hz, and high $H_z$ as presented using a quadratic ordinate. A straight line gives the $H_z^2$ terms while the parabolic contribution gives the $H_z^4$ terms.

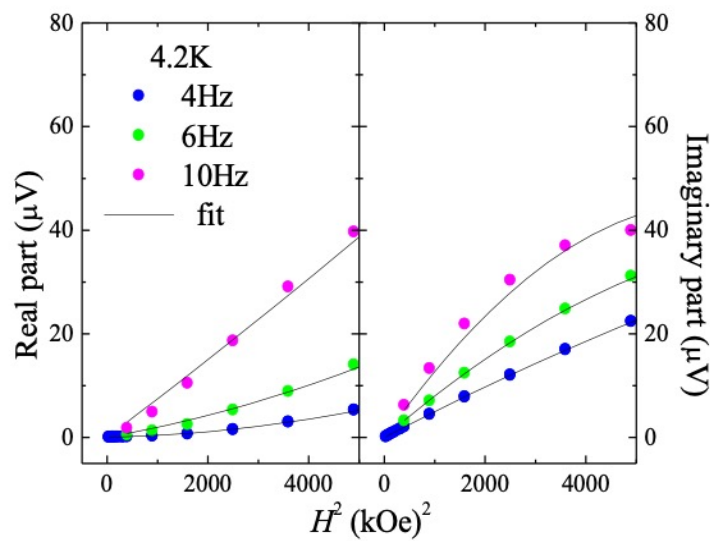

Figure 5, Longitudinal TME at various $f_d$, 4.2 K, and high $H_z$ as presented using a quadratic ordinate. Note that the induction of 10 Hz is abnormal at 90 kOe.

**Table 2.** High-$H_z$ (> 20 kOe) longitudinal TME impedance, which is decomposed into 5 components using $c_{r0} + (c_{r2} + ic_{i2})H_z^2 + (c_{r4} + ic_{i4})H_z^4$ with a 13-mA current drive, where pp$\Omega$ is $10^{-24}$ $\Omega$. We fixed the zero-field resistances of 4.5 μΩ and 27 μΩ in the parameter fitting.

|  | $c_{i2}$ (fΩOe$^{-2}$) | $c_{i4}$ (ppΩOe$^{-4}$) | $c_{r0}$ (μΩ) | $c_{r2}$ (fΩOe$^{-2}$) | $c_{r4}$ (ppΩOe$^{-4}$) |
|---|---|---|---|---|---|
| 4Hz @4.2K | 394±2 | -9.2±0.2 | 4.5 | 15±2 | 13.0±0.3 |
| 4Hz @10K | 400±2 | -9.2±0.2 | 4.5 | 14±2 | 13.3±0.3 |
| 4Hz @100K | 413±5 | -6.4±0.7 | 27 | 5±1 | 0.9±0.2 |
| 6Hz @4.2K | 652±6 | -35±1 | 4.5 | 130±10 | 15±2 |
| 10Hz @4.2K | 1190±30 | -110±5 | 4.5 | 560±30 | 6±5 |

Figure 5 illustrates the vortex dynamics for varying $f_d$. The cyclotron frequency $f_{LLL}$ increases from 0.8 to 4.5 Hz between 17 and 90 kOe. A drop of induction at 90 kOe, 10 Hz, and 4.2 K may reveal an interesting dynamic of the $S^8$ soliton for future study, i.e., 10 Hz exceeds twice $f_{LLL}$ ~ 4.5 Hz. We therefore removed this anomalous point before fitting the parameters. The $c_{i2}$ terms grow faster than $f_d$. The pure Eddy loss $c_{r4}$ should increase with quadratic frequency. This increased slightly at 6 Hz but surprisingly reduced at 10 Hz. Therefore, $c_{r4}$ could not be attributed to the Eddy loss alone, which contained the spectral flow, as mentioned in the preceding paragraph. We scanned the frequency response at 20 and 90 kOe. There was a resonant frequency that decreased slightly with $H_z$ between 20 and 30 Hz. The resonance is nearly independent of $H_z$ but becomes softer when $H_z$ increases from 20 to 90 kOe, which supports the STB mode, as discussed later on.

The spectral flow of $c_{r2}$ should be proportional to $f_d$, as induced by the spectral flow. Surprisingly, it increased anomalously more than 8-fold from 4 to 6 Hz. It is easy to estimate a quadratic curve using three $f_d$ points, leading to a negative $c_{r2}$, for instance at 3 Hz.

## 3. Discussion

### 3.1 The lotton physics

We have previously argued qualitatively that the γ-dopant acquired a micron size but was incapable of identifying the lotton [12-15]. BEC occurred when the estimated lotton densities of $^{103m}$Rh and $^{93m}$Nb exceeded $10^{12}$ cm$^{-3}$ at room temperature. At that stage, we did not know the reason why $^{93m}$Nb and $^{103m}$Rh had similar critical densities; now we have resolved this conundrum and report the $^{93m}$Nb quantum-critical density $1.70 \times 10^{12}$ cm$^{-3}$ as $\nu_{qc}$ = 0.88. Both kinds of p-wave lottons have similar $N_I$.

Most recently, either a low-threshold or even a thresholdless laser using the resonant topological insulators have been found theoretically and experimentally, employing the unidirectional γ propagation at the edge state [82-84]. Our topological γ laser is quite different. The matter-field coupling of $N_I α$ provides an off-resonant but strong-coupling γ blockade at ν >> 1 [15]. Incoherent γ-rays emitted from a radioactive source turn into two coherent beams of opposite helicities propagating on the surface states, which are probably entangled too. One-way propagation at the edge states between two different topological indices is the hallmark of topological photonics. The Sisyphus cycle reversed when the positions of the source and sample were exchanged. The theoretically restored $\mathcal{T}$- and $\mathcal{P}$- symmetries of the AF ladders are broken. We thus speculate that the cosmic axion also breaks the $\mathcal{PT}$-symmetry to give complex eigenvalues [4, 5].

For the past decade, x-ray monitoring has continued to indicate a $ν$-dependent half-life, i.e., < 16 years at $ν > 1$ whereas it is probably > 16 years at $ν < 1$. We provide some general discussion on the lotton physics for the TME and x-ray results, which are probably related to the detection of invisible particles in the cosmos.

*3.2 Gravitational effects*

A. X-ray signals

The Peierls transition lowers the 5p level by pairing the Mott electrons in the $^{93m}$Nb NKM groundstate. The Rh atom has a low-lying excitation, where the 5s electron transfers to the 4d orbit providing a 4d hole in the filled 4d shell. The 5p and 4d levels cross each other by increasing $B_\pm$, because the 5p level increases much faster than the 4d level in an exact 2D solution of the Mott exciton [85, 86]. However, the real spectrum in the absence of LLs lies between these two boundary conditions of 2D and 3D. The fact that 39.8 keV/(3×10²) ~ 130 eV but the vacuum splitting $\hbar ω_R$ ~ 100 eV in regime I of $ν < 0.3ν_{qc}$ [10] (see Appendix A) indicates d-wave mixing in NKM at 300 K. Accordingly, the Mollow triplet of $^{103m}$Rh in previous reports [10, 12] is only the central part of the mixing quintuplet spectrum, which leads to an imprecise triplet fitting of $\hbar ω_R$ in reference [10].

The Rabi gap $\hbar ω_R$ of the $^{103m}$Rh lotton opened from 100 to 500 eV before it underwent BEC over the interval $0.3ν_{qc} < ν < ν_{qc}$ in regime II [10]. The pressure squeezes lottons to reduce the Yukawa tail leading to a long list of descendent consequences, i.e., enhanced Zeeman splitting, the increasing $B_\pm$, $\Delta_P$ and $T_{Kl}$. The level crossing provides freedom to shift $\hbar ω_R$. After the initial $\hbar ω_R$ had opened by bremsstrahlung pumping at 300 K, the enhanced gap remained for the whole decay time, even at $ν \ll 0.3ν_{qc}$.

The initially open $\hbar ω_R$ in regime II shrank abruptly and then proceeded to collapse and revival [87], when lowering $T$ by LN₂. Figure 6 in our previous report [10] demonstrates only one half of the collapse-and-revival transition lasting for one hour. Before the transition terminated, we stopped the LN₂ cooling. The lowered $\hbar ω_R$ abruptly recovered its initial $\hbar ω_R$, when $T$ slowly returned to

300 K at $\nu \ll 0.3\nu_{qc}$ [10]. The $^{103m}$Rh octonions always remember the initial NKM mixing whenever the bremsstrahlung pumping stops, no matter how we change the *T* settings during the decay.

We previously explained this peculiarity by referring to condensed lotton droplets [14] to preserve the initial state. This fails to explain the open Rabi gap. We understand now that lottons remember their birth twists. The strong $B_\pm$ field mixes the NKM orbits, as revealed by the enhanced pumping efficiency in regime II [10]. The initial mixing is then preserved over the decay time for $\nu \ll 0.3\nu_{qc}$.

Pumping $^{103m}$Rh at $\nu > \nu_{qc}$, the lotton did not immediately undergo BEC in regime III at 300 K (see Appendix A). Instead, $\hbar\omega_R$ switched twice abruptly between two levels of 100 and 200 eV before BEC, which has been recognised as collapse and revival [12]. After completing the collapse-and-revival transition at 300 K, the condensate returned to vacuum splitting but at $\nu \sim 0.5\nu_{qc}$, the excitation number of which had dropped into regime II. The $^{103m}$Rh lottons remembered the initial ν in regime III again. The heavy d-wave hole reduces $N_l$ according to (1), which releases the pressure between lottons. Hence, we suggest a d-wave lotton BEC in regime III.

The LN$_2$-cooling response of $^{103m}$Rh is quite complicated, and depends on the initial ν, the *T* settings, and the decay time [10]. Further investigation is clearly required to give a more comprehensive picture. As mentioned in the preceding section, lowering *T* by LN$_2$ opened $\hbar\omega_R > 100$ eV in regime I. The $\hbar\omega_R$-opening response to cool p-wave lottons in regime I is at odds with the $\hbar\omega_R$-shrinking response to cool d-wave lottons in regime II [10]. The standalone anapole offers new insights for resolving this puzzle, which has remained unsolved for more than a decade.

On cooling the BEC $^{103m}$Rh at the initial $\nu > \nu_{qc}$, slow breathing modes of $\hbar\omega_R(t)$ emerged spontaneously (see Fig. A4 in Appendix A), which caused a persistent slow-varying amplitude between 100 and 500 eV along with LN$_2$ cooling over the whole decay time when $\nu < \nu_{qc}$. We repeated the cooling experiments four times in a series of one-day studies, using different cooling periods. Pronounced peaks of the breathing $\omega_R(t)$ spectrum in the mHz region are better identified in the longest cooling record for almost three hours [11]. We focus on the well separated peaks > 1 mHz due to the limited resolution. It is interesting to compare the six major peaks in Fig. A4 of Appendix A with the GW frequencies $f_{GW}$ of known strong sources [88].

We have long suggested one possible scheme to detect GWs by means of gravitational Raman effects [11, 89]. A $p_i \overleftrightarrow{h}_{ij} p_j$ perturbation deforms the $^{103m}$Rh nuclear shape to accelerate the forbidden E3 transition, which yields a contribution of the allowed E1 transition, where $p_i$ is the momentum of the hadron field and $\overleftrightarrow{h}_{ij}$ is the long-wavelength spin-2 tensor field [89]. The nuclear structure is very hard, and its shape seems hard to deform. We thus express the GW action in an alternative way. According to general relativity, $\overleftrightarrow{h}_{ij}$ deforms free space, where the nuclei live. The emergent condensate in LN$_2$ with the initial $\nu > \nu_{qc}$ causes a persistent edge quadrupole zero mode of AF d-wave ladders spatially and a QTC Goldstone mode temporally. Hence, the spin-2 edge state in the absence of any EM Floquet drive is different from the spin-3/2 MZM, as described in the next section. The GW Floquet drives of $\vec{h}_{ij}$ mix the E3 and E1 periodic protocols of the QTC quantum walks in rotating reference frames, which tune $\theta_L$, $B_\pm$, and thus $\hbar\omega_R(t)$, simply because the E1 polarization of a nontrivial chain corresponds to $B_\pm$. Whether or not this is an anapole condensate requires more detailed investigation.

B. TME signals

According to the gap-opening restriction $hN_I f_d > \Delta_S(T_{on1})$, raising $T > T_{on1}$ opens the domain wall in the presence of twist memory. The Floquet drive using the incommensurate $f_d = 3.3$ Hz cannot open any domain wall at the earth's field (see Fig. B1 in Appendix B), because $|2f_d - 4 \times 92 f_{LLL}| > 1/\tau_{R1}$. Only the near-commensurate $f_d = 6$ Hz with $m_H = n_H = 1$ installs the twist memory of octonions when lowering $T < T_c$, which reopens one and only one domain wall by raising $T > T_{on1} \sim 8.1$ K in the first test. We raise accordingly several questions about the twist memory. Can an arbitrary $f_d$ create an edge state at an arbitrary $T_{on1}$ satisfying $hN_I f_d > \Delta_S(T_{on1})$ after installing the twist memory using 6 Hz? Can $f_d$ create a complicated domain-wall structure instead of a single domain wall at the BPS $T_{on1}$? Can we apply $H_z$ to remove the gap-opening restriction by mixing the MRP and PHP flavours?

We now use some observations to explore further the mysterious twist memory, which is caused by the residual $H_z \gg$ the earth's field in the second test. The near-commensurate $f_d = 10$ Hz implants the twist memory into the $\mathcal{T}$-breaking intermediate state. The TME terminates abruptly by lowering $T$ from $T_c$ to 4.2 K, when the residual $H_z < H_{c1}(T)$. The top-down capping Abrikosov lattice in Fig. 1e at 9 K dissolves to approach the BPS condition at 4.2 K. We assume the PHP-magnet is independent of the weak $H_z$, i.e., an $H_z^0$-magnet with a nonzero $C_1$. The Floquet drive using $f_d = 2$ Hz recreates an intermediate state in the presence of a pronounced third harmonic, however, which is the near-commensurate $f_d = 6$ Hz of the MRP flavour at zero $H$. This implies that the zero-field twist memory is retained if we remove the residual $H_z$. Does the 2-Hz edge state mix the MRP and PHP flavours? Regardless of what happens at 2 Hz, the complex PHP flavour removes the commensurate and gap-opening restrictions for all drive frequencies at 4.2 K, which provides the emergent $H_z^0$-magnet except at the exact commensuration.

The twist memory is nothing but the spacetime texture, which is most likely independent of the octonion flavours. For example, the condensate preserves the large $n_L$ at the moment of changing octonion flavours, as revealed by the anomalous topological Kondo effect. The QTC twist memory breaks spontaneously the continuous $\mathcal{T}$- and flavour symmetries in a thermal equilibrium. The inductance $\propto T > 72$ K in Fig. B3 in Appendix B is attributed to the kinetic energy of gyrating

$\mathbb{T}^4 \times S^4$ anapoles. An equivalent Kopnin mass $\propto T$ depends on the electronic screening, which tunes a breathing frequency $f_B(T)$ as a function of $T$. Two resistive $T$-plateaux and the sag are caused by the near and exact commensurations among three periodic drives of $\omega_\gamma$, $f_d$, and $m_a$, respectively. We postpone our discussion of the axion coupling to the next section while we focus first on the GW interaction.

The restoring PH-symmetry reduces the Kopnin mass by half. Hence, the shrinking PHP $f_{LLL}$ ~ 13.217 mHz is caused by a spontaneous PHP $B_z$ ~ 6.771 mT. The $H_z^2$- and $H_z^0$-supermagnets refer to the vortex dynamics of distinct Chern classes, which are characterized by the zero and nonzero $C_1$, respectively. We observed several kinds of nontrivial domain-wall structures, which allow us pick up the voltage in the presence of Cooper-pair moat. The interplay between PHP and Abrikosov vortices provides the first kind from the BPS PHP $H_z^0$-supermagnets. The BPS MRP $H_z^2$-supermagnets provides the second kind at 4.2 K, while the inhomogeneous $T$ distribution provides the third kind for the quantum anomalous $T$-plateau of RRP vortices at $T > T_c$.

We apply the exact commensuration to select the GW signals in the second test. The spin-1 drive is dressed on the NKM pseudospin to provide a PHP cat state at 4.2 K, which is the seesaw quadrupole zero modes of $Q_z cos(f_d t)$. Figure 1d shows a schematic picture of the seesaw quadrupole $Q_y cos(f_d t)$ at the earth's field in the first test, which is rotated to $Q_z cos(f_d t)$ by the residual $H_z$ in the second test. Although the MRP $Q_y cos(f_d t)$ and the PHP $Q_z cos(f_d t)$ have distinct flavours, the flipping of the two MZM quadrupoles provides the similar $V_y cos(f_d t) cos(f_{GW} t)$ on the pickups. However, the trivial conductivity of solders and wires in via holes provides a $T$-dependent $I_x$ distribution, which causes the sawtooth function in the second test.

We list the following eight reasons to explain the mHz GW signals shown in Fig. B4 in Appendix B.

1) The GW signals are caused by MZM quadrupole flipping.

2) We tuned $f_d$ to select the GW signals, which are always there.

3) The GW signals are repeatable by proper selection of $f_d$, and are not coincidental.

4) The GW signals are quantized, caused by the flipping of quantum edge states.

5) The GW signals switch abruptly one-by-one in time, caused by the earth's rotation.

6) The GW signals are not continuously distributed, caused by a limited selection of available GW sources.

7) The GW signal is independent of superconductivity, and is not caused by EM waves.

8) The most profound evidence is the topological GW signal evolving abruptly backwards in time whilst losing superconductivity. Irrespective of the reason, the mHz signal is not caused by any error of instrumentation. More precisely, the lotton flavour transition changes the gravitating Majorana mass.

Unfortunately, despite the data accumulated to date, we are far from being able to identify the four-fold symmetry in 24 hours to detect GWs.

When $f_{GW} = \pm 2f_{LLL}$ matches the Majorana mass, passing GWs $\overleftrightarrow{h}_{xy}cos(f_{GW}t)$ along the z-axis flips the LLL pseudospin of $Q_z cos(f_d t)$ by means of the $\langle m'|\boldsymbol{\sigma} \cdot \nabla \times \overleftrightarrow{h}_{xy}\boldsymbol{P}|m\rangle$ term [89], where the magnetic quanta are $m'$ =1 and $m$ = -1 and vice versa, $\boldsymbol{\sigma}$ is the LLL pseudospin, and $\boldsymbol{P}$ is the momentum of coherent matter fields. Two indices z and y switch with each other between the first and second tests. Recent work also reveals that both the electric and magnetic fields flip the multipolar nuclear spin [90].

How can $f_{GW} \ll 2f_{LLL}$ flip the MZM quadrupole? The nuclear magnetic resonance does not require an exact frequency to flip the nuclei, and depends on a quadrupole relaxation time $\tau_{R2} \gg \tau_{R1}$. The near-commensurate $f_d$ creates the dipole edge cat states, as long as $|m_H f_d - 4n_H \times 92 f_{LLL}| < 1/\tau_{R1}$ (see Fig. B1 in Appendix B). However, $f_d$ = 6 Hz yields no GW signal in the first test, because $|m_H f_d - f_0| > 1/\tau_{R2}$.

We previously reported a GW signal in reference [11]. Three values of $f_d$ of 3.3 Hz, 6 Hz, and 22 Hz are presented in the study of quantum anomalous plateaux, (see Fig. B1 in Appendix B). We measured the GW signal using $f_d$ = 1.1 Hz @ 4.3 K at earth's field for half an hour, because the TME contains an oscillation of two sinusoidal components, i.e., a strong $10f_{GW}$ signal ~ 63 ± 1 mHz with an amplitude of about ± 2 µΩ and a weak $f_{GW}$ signal of about ± 0.2 µΩ. We assume this is a GW signal with $m_H$ = 5, $n_H$ = 1, and a Lagrange multiplier $n_G$ = -10. A new matching condition of

$$|m_H f_d - 4n_H \times 92 f_{LLL} - n_G(2f_{LLL} - f_{GW})| < 1/\tau_{R2} \tag{5}$$

is provided by three periodic drives of $\omega_\gamma$, $f_d$, and $f_{GW}$, where $\tau_{R2}$ is about 300 seconds, as shown below. We thus yield the only solution of $4 \times 92 f_{LLL}$ ~ 5.750 Hz ± 3 mHz and $f_{LLL}$ ~ 15.625 mHz ± 8 µHz for the first test. The weak $f_{GW}$ signal is probably caused by a phase modulation, however, which helps us to identify the interacting GW. These complicated GW signals survived only at $T < T_c$.

We achieved the exact $f_d$ to match (5) in the second test, which allowed us to measure the GWs freely by tuning $f_d$. We simplify (5) by

$$f_0 = 4n_H \times 92 f_{LLL} + 2n_G f_{LLL} \tag{6}$$

for the GW signals in Fig. B4 in Appendix B, where $n_H$ = 1, $m_H$ = 1, $n_G$ = 1, $f_0 \cong 4.8903$ Hz and $f_{LLL}$ = 13.217 mHz. The positive $n_G$ sign will be explained later on. To select one GW arriving at earth depends on $|f_d - f_0|$, the strain amplitude, and the polarization. For example, $f_d - f_0$ = 2.7 mHz using $f_d$ = 4.893 Hz picked up two GWs of 1.5 and 3.6 mHz one by one, which happened to be more suitable than any other GWs available near 2.7 mHz at that moment [88]. The earth's rotation changed their privileges, which switched spontaneously from 1.5 to 3.6 mHz after two probably quantized cycles. To verify this argument, we restored the extinguished signal at 1.5 mHz by a new matching condition using $f_d - f_0$ = -2.3 mHz half an hour later. Hence, we adopt $f_0$ = 4.8903 Hz ± 0.2 mHz. The Q-value of γ-condensate guarantees the precision as an optical clock [33, 34]. The signal at 1.5 mHz propagated backwards (see the left panel of Fig. B4 in Appendix B), when we restored it by changing $f_d - f_0$ > 0 to $f_d - f_0$ < 0.

The TME using $f_d$ = 4.889 Hz caught the GW signal of 0.53 mHz with $f_d - f_0$ = -1.3 mHz < 0 (see the right panel of Fig. B4 in Appendix B). When we raised $T > T_c$, the GW signal remained at the vanishing superconductivity. The signal evolves backwards in time, when $f_d - f_0$ + 13.217 mHz > 0 changes sign by advancing $n_H$ in (6) from 1 to 2, because the PHP $f_{LLL}$ is cut by half. The presence of Eddy loss reduces $\tau_{R2}$ for instance from 300 to 50 seconds to satisfy (5), which has been revealed by the quantum anomalous $T$-plateau of RRP vortices. This is the reason for assuming a positive $n_G$ = 1 for $f_0$ = 4.8903 Hz. The persisting $n_G$ = 1 also reveals a selection rule owing to the global PHP Berry phase in (6), which picks up an odd $n_G$ in the second test and an even $n_G$ in the first test. No GW signal was yielded using $f_d$ = 2.448 Hz at 4.2 K, which provides a lower bound of $\tau_{R2}$ > 175 seconds. We thus assume 300 seconds for $\tau_{R2}$ at 4.2 K in the two tests.

We measured $V_y$ across the pickups of v₃ and v₄, which is the same setup used to measure the quantum anomalous $T$-plateaux. This is the reason for the constant sawtooth amplitudes, which are independent of the strain amplitudes of GWs arriving at earth. The sawtooth amplitudes ~ ± 0.4 μΩ are smaller than the quantum anomalous $T$-plateau ~ 1.5μΩ, which is caused by the wiring. However, AF ladders are not perpendicular to the voltage pickups, which allows us to sense a small $V_y$. Had we observed the transverse term (see Fig. B2 in Appendix B), the signal would have been much larger than ± 0.4 μΩ.

*3.3 The cosmic dark matter*

We observed the absolute negative resistance to gaining energy many times at ν > 1. One of these, the reading 0.33 μV with a phase of -102° at 9 T @ 10 Hz, and @ 4.2 K is shown in Table 4 in our previous report [12]. We have long speculated coherent axion coupling [68] in the presence of annular ferromagnetic polarizations from the gofour anapoles, which causes the negative resistance in the anomalous topological Kondo effect at the residual $H_z$ (see Fig. B3 in Appendix B). The thermal potential is necessary to release the Floquet-Kondo screening in anapoles. As a consequence, two Floquet drives of $I_x$ and $a_{DM}$ together transform the soliton flavour by creating the topological line defect to conduct $I_x$. AF anapoles are caught by the annular field of $I_x$ to condense. On the sag tip at 90.5 K, the anapole MZM condensate provides a commensurate signal from the cosmic axion $a_{DM}$.

The time derivative of $a_{DM}$ drives the opposite chiral magnetic currents $I_{CME} = N_I g_{a\gamma\gamma}\sqrt{2\rho_{DM}}B_+$ of nuclei provided by an axion coupling of $g_{a\gamma\gamma} \sim 10^{-14}$ GeV$^{-1}$ [78] along the annular $\pm B_+$ of the clockwise and counterclockwise anapoles. Two Floquet currents $I_{CME}$ and $I_x$ introduce a net mm-wave magnetic field $\propto sin(f_d t) \times sin(m_a t)$ that penetrates the anapole dimer chain. The chain minigap levels are free from the factor $N_I$ but yield a triplet structure, i.e., two side bands $\pm 2hf_d$ sandwiching the main gap $\Delta_a = 2hm_a$. The nontrivial anapole dimer chain provides a charge- and pseudospin-density wave along the x-axis, which is perpendicular to the instanton dimer rings. Two quantum anomalous $T$-plateaux preserving the flavour symmetry validate the incompressible anapole liquid, and contain two Su-Schrieffer-Heeger chains in series, i.e., one ring-type chain and one linear chain. More precisely, the resistance 6.38μΩ ≈ $8\nu R_K/N_I$ between plateaux reveals the 8-fold degeneracy of paired octonions, which are locked on the plateaux by the commensuration

$$|m_H N_I f_d - 4n_H N_I \times 92 f_B(T) - n_A m_a| < 1/\tau_{R0}, \qquad (7)$$

where $f_B(T)$ is the $T$-dependent MZM frequency tuned by the active levels, $n_A$ is an axion Lagrange multiplier, and $\tau_{R0} \gg N_I \tau_{R1}$ is a spin-0 relaxation time constant. Two $T$-plateaux are provided by distinct $m_H$ and $n_H$ because $\tau_{R1} > N_I/m_a$ provided by $n_A = 1$ in (7), where $n_A m_a$ is irrelevant for two periodic drives of $\omega_\gamma$ and $f_d$. The sag tip corresponds to an exact commensuration for three periodic drives of $\omega_\gamma$, $f_d$, and $m_a$, the $T$-plateau of which is smeared by the thermal potential. Had we reduced $f_d$, a deeper sag of the same width would have been moved accordingly to a $T < 90.5$ K.

The V-shaped negative resistance (see Fig. B3 in Appendix B) is caused by a product of two $T$-functions, going up and down with the inductance. Hence, the minimal resistance is located roughly on the half-way point towards the loss of inductance. When raising $T$, $I_x$ withdraws the persistent axion conversion [78] increasingly from the nuclear heat reservoir, but the electronic screening fades out to transfer the axion power into the electronic reservoir. The cascade chiral anomalies of non-Hermitian terms give rise to a slow $f_d$ component of $E_x \propto I_x(t) N_I^2 \partial_x a_{DM} \dot{a}_{DM} B_+^2$ rather than a fast $m_a$ component of $E_+ \propto \partial a_{DM} B_+$ [1], where the axion wind blows a couple of nW into the 0.1-cm³ working sample. The resistance to gaining energy is attributed to the negative spectral flow, i.e., the $f_d$-breathing net $I_{CME}$ along anapole rings plays the role of anomalous Hall current to polarize the anapole dimer chain. A gapless mode arises from the tunnelling between the anapole dimers gapped by $\Delta_a$, and gives rise to the sag on the negative quantum anomalous $T$-plateau at 90.5 K (see Fig. B3 in Appendix B). The sag width of 1.8 ± 0.2 K may narrow down the uncertainty of $m_a$ [76] to be 78 ± 9 μeV. The most recent ADMX reports [91, 92] support this result.

Figure B3 in Appendix B shows the stable anapole at 100 K, which may extend to $T \gg 300$ K in the presence of a commensurate Floquet drive, as visualized by magnetic powders [11]. We may thus have MZM anapoles $\gg 300$ K for a quite long time at zero $H$. It would be better to search directly for the $m_a$ frequency [76-79]. Unfortunately, we did not pay sufficient attention to the 19-GHz wave emitted from the Primakoff conversion, which is detectable in the ADMX-HF experiment [93].

We now switch to the possible negative spectral flow of $H_z^2$-resistance with $f_d < 3$ Hz in Fig. 5. For increasing $H_z$, the competition between the positive $H_z^2$-inductance and the negative $H_z^4$-inductance changes the TME response from a magnetoinductance to a magnetocapacitance, where the amplitude drops and the phase rotates from π/2 to -π/2 when passing a turning point of $H_z$ depending on $f_d$. The $H_z^2$-inductance is linear in $f_d$ but the $H_z^4$-inductance induced by the Eddy current grows with $f_d^2$. It imitates a resonance, the frequency of which reduces as a linear function of $H_z$. However, we observed a TME resonant frequency around 20 Hz at ν ~ ν$_c$, which is almost independent of $H_z$. The resonance became soft when $H_z$ was increased from 20 to 90 kOe. The resonant frequency of the STB Goldstone mode depends weakly on $H_z$ while the thinner vortex at higher $H_z$ gives rise to a lower Q.

Put simply, the buckled RRP vortex looks more or less like a dimer chain of breathing anapoles. The negative spectral flow becomes more apparent either by lowering $f_d$ to reduce the Eddy loss or by rotating the RRP vortices to the cm-long axis to increase the buckling amplitude, as revealed by a measurement of the large-$n_L$ RRP vortices in the fourth test (see Fig. B2 in Appendix B). We might have obtained interesting results in terms of a gain in energy in the non-Hermitian photonics [4, 5] many times.

**Yao Cheng:** Experimental and theoretical works, Writing **Ben-Li Young:** Low-$T$ experiments.

**Acknowledgements**

We are indebted to Alex Chao from Stanford University for his encouragement using the 6-MeV linac to activate $^{103m}$Rh at an early stage of our long-term study. We would also like to thank Chih-Hao Lee from the national Tsing Hua University, Hsinchu, for donating and activating two single crystal Nb samples; Yuan-long Liu for the second and third experiments in Chiao Tung University, Hsinchu, Hong-Jian He from Tsinghua University, Beijing, for valuable discussions on the quantum field theory.

# APPENDIX A: Review of the $^{103m}$Rh x-ray experiments

*A.1 Introduction*

The 30-keV γ emitted from $^{93m}$Nb is too rare to be observed due to the extremely large internal conversion rate [63]. In contrast, the 39.8-keV γ emitted from $^{103m}$Rh is directly observable. We activated the pure Rh sheet using a 6-MeV bremsstrahlung source [10]. When the $^{103m}$Rh densities

$D_{Rh} < 3 \times 10^{11}$ cm$^{-3}$, the Rabi gap $\hbar\omega_R$ stayed at ~ 100 eV, which was recognised as vacuum Rabi splitting in the so-called regime I at 300 K [10]. We have discussed in the main text that the use of $\hbar\omega_R$ to describe the Mollow triplet is imprecise, and it is indeed a mixing of the p-wave triplet and the d-wave quintuplet. The previously estimated $D_{Rh}$ is lower than the true density, because the γ counts were taken along the shortest axis. We now know that a strong $B_I$ leads to more γ counts along the longest axis. To avoid confusion, we still apply the old $D_{Rh}$ hereafter as previously reported [10-15].

The Rabi gap $\hbar\omega_R$ increased from 100 eV to 500 eV when $3 \times 10^{11}$ cm$^{-3}$ < $D_{Rh}$ < $10^{12}$ cm$^{-3}$, where the pumping efficiency increased with $D_{Rh}$ in regime II [10]. When $D_{Rh} > 10^{12}$ cm$^{-3}$, it entered regime III. The Rabi gap changed abruptly every 20 minutes for one hour, which is now identified as BEC "collapse and revival". After that, the Rabi gap maintained the vacuum splitting ~ 100 eV for the remainder of the decay time. Interestingly, $D_{Rh}$ stopped increasing, even with further pumping in regime III. We speculate a quantum-phase transition $\nu_{qc} = 1$ for $D_{Rh} \sim 10^{12}$ cm$^{-3}$ as for the $^{93m}$Nb case, beyond which the decay was significantly enhanced, at least during the bremsstrahlung irradiation.

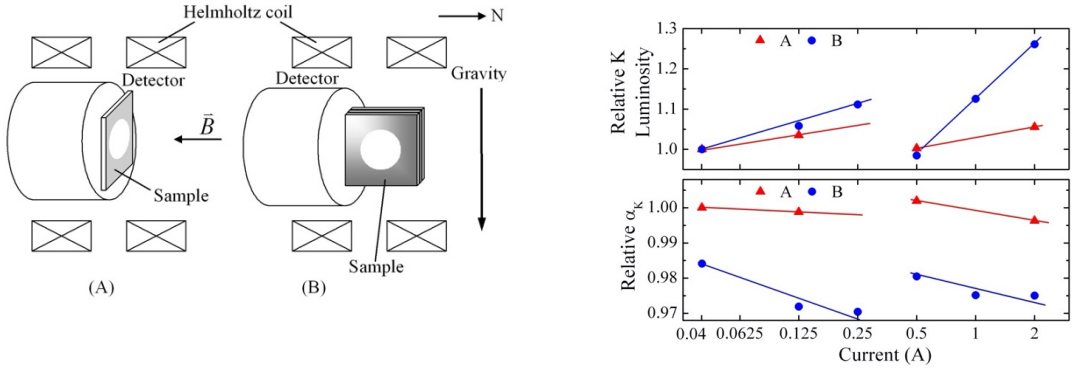

Figure A1, *Left*) Setups to show broken rotational symmetry of the $^{103m}$Rh lottons. The Rh sample sheet is a 25 mm × 25 mm × 1 mm sheet normal to the HPGe detector in configuration A. For configuration B, three rhodium samples are stacked together with two 0.05-mm plastic spacers. Two 0.5-mm tungsten sheets sandwich the package, where the long 25-mm axis is aligned towards the detector. The white spot indicates the position of the Bremsstrahlung irradiation. *Right*) Broken rotational symmetry, as revealed by the field-dependent luminosities of the 40-keV γ and the 20-keV K x-rays depending on configuration. The Helmholtz coil provides the horizontal *H* field of 7.62±0.01 Oe/A. The applied field $\boldsymbol{B} = \mu_0(\boldsymbol{H} + \boldsymbol{H}_E)$ contains the earth's field $\boldsymbol{H}_E$. The coefficient $\alpha_K \sim 110$ is the ratio of the x-ray and the γ counts. We set the luminosity at 0.04 A to be unity. The other luminosity is relative to this base level.

*A.2 Broken rotational symmetry and reorientation*

We now focus on the broken rotational symmetry of the $^{103m}$Rh AF ladders. Although the measurements were made in regime III, we believe the rotational symmetry was also broken in regime II. The orientation of the AF ladders was very sensitive to the applied field in terms of the Oe order. Figure A1 illustrates the measuring setups of two distinct configurations. The coil current of 0.04 A zeroes the horizontal earth's field $\boldsymbol{H}_{E\parallel}$, whilst leaving the vertical earth's field $\boldsymbol{H}_{E\perp}$ ~

0.25 Oe. The photoelectric attenuations of the 20-keV Kα x-rays and the 40-keV γ-ray are nearly equal. The right panel of Figure A1 shows the total counts of the K x-rays and γ-ray accumulated in three hours, the ratio of which is presented by the measured K-shell internal conversion coefficient $\alpha_K \sim 110$. The field dependence in configuration A is insignificant. In contrast, the field dependence in configuration B is more pronounced.

A critical field ~ 3 Oe abruptly reorientates the AF ladders from the horizontal direction to the vertical direction in configuration B, where the luminosity of K x-rays shows a logarithmic response to the applied field before and after the reorientation. The strong $B_I$ emanating from the nucleus also aligns all the atomic orbits in filled shells due to their non-spherical shapes. We thus observed a superradiance of K x-rays in Fig. A1 (right upper panel), which depends on orientation and field. K-x-ray superradiance has been reported for $^{93m}$Nb [14]. Therefore, broken rotational symmetry occurs at both nuclear and atomic levels. The logarithmic luminosity indicates the anisotropic lotton Meissner effect.

We could barely identify the field dependence of $\alpha_K$ in configuration B, although the error bars (not shown) are ~ 0.8 %, as evaluated using normal photon statistics. The error bars for the data points in the upper figure are much smaller, i.e., < 0.1 %. The 40-keV γ has a slightly stronger penetration than the Kα x-rays, which gives the 2% $\alpha_K$ discrepancy between configurations A and B. The reorientation of the lotton chains is clearly revealed by the field dependence of $\alpha_K$ in configuration B rather than the flat field response in configuration A (see the right-hand lower panel of Fig. A1).

*A.3 Standing waves of entangled γγ-rays*

The nuclear resonant absorption of $^{103m}$Rh employs at least two entangled γγ rays. Each γ of the entangled γγ has a continuous spectrum while their sum remains at 39.8 keV. Each of the γγ arriving at the detector has energy < 39.8 keV. The lower left-hand panel of Fig. A2 shows the ratio of the spectra of two different configurations in Fig. A1, which eliminates the detector profile to provide three peaks. The central peak o-ii is located at one half 39.8 keV. Two peaks of o-i and o-iii reveal the standing waves of entangled γγ matching the Rh lattice constant. The right-hand panel in Fig. A2 provides further evidence of entangled γγ. The detector cannot distinguish between γ and γγ, when two entangled γγ arrive simultaneously at the detector. However, the low-energy γγ can be removed by a filter. The x-ray counts located between 24 and 38 keV are the so-called Compton plateau of γ at 39.8 keV, which indicate photons scattered inelastically from the detector boundary. Part of the entangled γγ were removed by the filters, depending on their attenuations. We applied the same method to other radioactive sources without entangled γγ (not shown), which indicates identical Compton plateaux regardless of which filters were inserted.

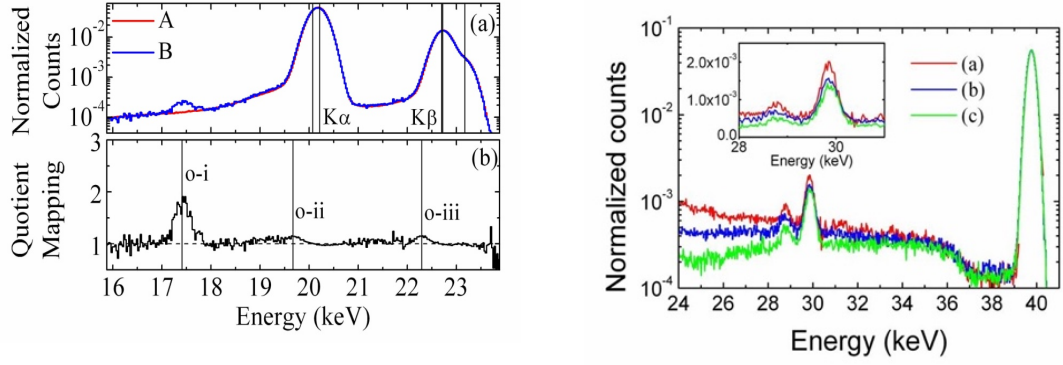

Figure A2, *Left*) The standing waves of entangled γγ. The upper figure (a) shows the K x-rays of $^{103m}$Rh, where the colours represent the configurations A and B in Fig. A1. An anomalous peak ~ 17.5 keV emerges in configuration B. A quotient mapping B/A in the lower figure (b) shows three peaks. *Right*) Evidence of the entangled γγ emission. We inserted filters to attenuate the x rays arriving at the detector in configuration A, i.e., no filter (red), one Cu sheet of a 35-μm thickness (blue) and one Ta sheet of a 25-μm thickness (green). The three curves are normalized with γ counts at 39.8 keV. The inset shows the γ escape peaks in the HPGe detector.

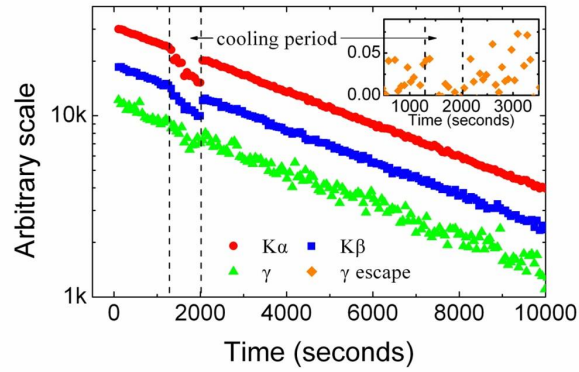

Figure A3, Cooling response of $^{103m}$Rh lottons with $D_{Rh}$ in regime I. The measuring conditions are identical to those for the A configuration as shown in Fig. A1. The γ escape is the escape peak of the 40-keV γ in the HPGe detector.

*A.4 Formation of the standalone anapole*

We quenched the Rh sheet using LN$_2$ for about 10 minutes in the A configuration (see Fig. A1). The initial ν was less than 0.3, i.e., in regime I. During the cooling period, the K x-ray spectra shifted abruptly to a higher level about ~ 100 eV (not shown). The spectral shift looks like something from an increasing Kα$_1$ along with a decreasing Kα$_2$. After the cooling period, the K x-ray spectra recovered abruptly while $\hbar\omega_R$ opened slightly to a value > 100 eV, i.e., slightly greater than the vacuum splitting. It is obvious that the linear drops in intensity seen in Fig. A3 are not caused by the absorption of LN$_2$, simply because both the 20-keV x-rays and the 40-keV γ are nearly transparent to the thin LN$_2$ film between the detector and the sample. The drop in γ escape counts in the HPGe detector also reveals the true reduction of 40-keV γs. More precisely, the LN$_2$ photoelectric attenuations are significantly different between the 20-keV x-rays and the 40-keV γ ray, and cannot provide the same degree of reduction. The abrupt recovering of intensities reveals that the linear reduction in time is not attributed to the LN$_2$ absorption; it is attributed instead to the detached boojums of anapoles.

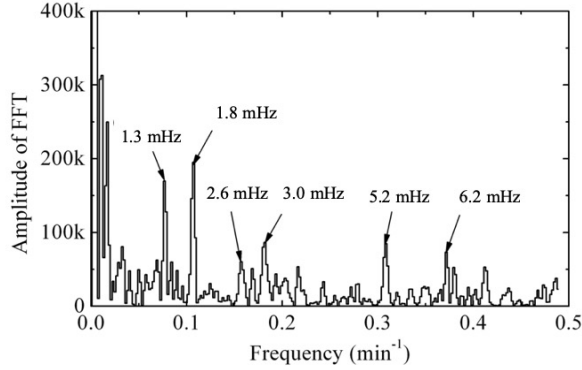

Figure A4, Spectrum of $\hbar\omega_R$ taken in the morning of May 24$^{th}$ 2006 in Beijing. The initial $D_{Rh}$ was pumped in regime III. The measuring setup was the same as configuration A in Fig. A1 but was cooled by LN$_2$.

*A.5 Slow breathing mode of the Rabi gap*

When we cooled the Rh sheet with the initial ν > 1, $\hbar\omega_R(t)$ oscillated slowly (see Fig. A4). The sampling rate of x-ray spectra was one minute plus a couple of seconds caused by the software. The sampling duration lasted almost 3 hours, where the $D_{Rh}$ decayed about 10 times. However, the breathing $\hbar\omega_R(t)$ remained consistent throughout the decay time.

# APPENDIX B: Review of the $^{93m}$Nb TME experiments

We carried out four sets of TME experiments, as detailed in the main text. The first was opened in arXiv:1102.1766, the second and the third were described in [12], and the fourth was partially introduced in [11,15].

*B.1 Longitudinal quantum anomalous plateaux in the first test*

As soon as we realised that the γ-dopant of $^{103m}$Rh acquires a mass from the longitudinal phonon, the first TME experiment of $^{93m}$Nb was undertaken to investigate the interaction of the Cooper pair also with the phonon. We applied a cold head to cool samples in Beijing, where the sample y-axis is aligned along the vertical direction while the x-axis is along the NS direction. The heat pump worked with a frequency near 1 Hz, which caused a measured temperature fluctuation of ~ 0.1 K on the cold head.

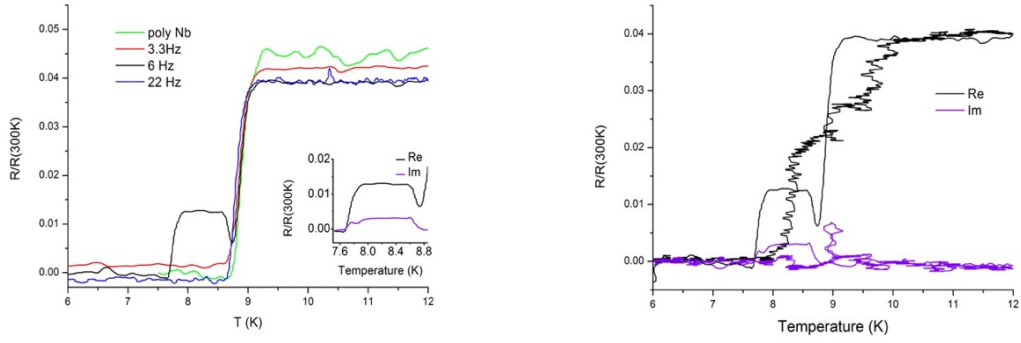

Figure B1, *Left*) TME impedance of pure Nb samples. One sample is the activated single crystal and the other is a dirty poly-crystal. We show the ratios to the room-temperature resistance noted by R(*T*)/R(300), which are independent of sample type. It is clear that the activated single crystal has an overly large residual resistance. The residual resistance is 4.5 μΩ at the ratio R(10)/R(300) = 0.04. R(10)/R(300) of a single-crystal Nb usually lies in a range from 0.02 to 0.01. We have elevated the red line for clarity. *Right*) Hysteresis between cooling and heating cycles. The wiggling signal is caused by the heat pump, which demonstrates the phase of cooling.

We noted a systematic *T* hysteresis ~ 1 K between the heating and the cooling procedures, as caused by the delay in heat transfer between the cold head and the sample. The search for the common phonon interaction between two condensates led us to fix the $T_c$ in the figures by hand at 9 K in order not to miss any possible shift in $T_c$ at that time. $T_c$ was actually 9.3 K rather than 9 K, as verified by careful measurements made many times in Hsinchu half a year later. The underlying physics do not matter. Therefore, we retain the *T* abscissa of the originally presented figures hereafter to avoid any confusion. However, we need to bear in mind that the reading of *T* must be shifted by 0.3 K. For example, the TME onset temperature $T_{on1}$ is 7.8 K here in Fig. B1 but 8.1 K in the main text.

Two longitudinal quantum anomalous *T*-plateaux ~ 1.5 μΩ and ~ 3.1 μΩ emerged along the longitudinal direction for 7.8 K < *T* < 8.6 K and 9 K < *T* < 9.5 K, which showed up only at a particular frequency of 6 Hz ranging from 1 to 40 Hz (see Fig. B1). The drive current was 22 mA. One of the plateaux was slightly inductive, which implies that the true commensurate frequency 5.750 Hz is below 6 Hz. Figure B1 clearly demonstrates the Lifshitz transition to form a domain wall depending on temperature. The temperature fluctuation at the cold head provided the wiggling behaviour induced by the 1-Hz heat pump, which was absent during the heating cycle.

*B.2 Hysteresis of the rotational TME map in the fourth test*

We measured the hysteresis of the transverse terms at 4.2 K (see Fig. B2), which relates to the TME response between $v_5$ and $v_6$ along the y-axis, as shown in Fig. 1c. Rotating the sample in the dewar yielded the rotational TME map in the fourth test. We omit here the longitudinal terms to be included in a future report, which show more complicated results regarding *T*, *H*, $f_d$, $I_x$, $\theta_R$, and the operational paths at 2 > ν > 1. Some results have been previously discussed as in Figure 2 in reference [11], which are reassessed to make the whole picture clearer. Figure B2 illustrates the results using $I_x$ = 200 mA, $f_d$ = 20 Hz, and *H* = 10 kOe. The transverse resistance contained significant higher harmonics and depended on *H*. Eight driving parameters were taken one after another at each angle $\theta_R$ setting, e.g., 20 Hz, 4 Hz, 25 mA, 50 mA, 100 mA and 200 mA, which provided similar but distinct results regarding the rotational angles $\theta_R$. The stepwise resistive increments regarding some particular $I_x$ currents are not simply attributed to the nonlinearity.

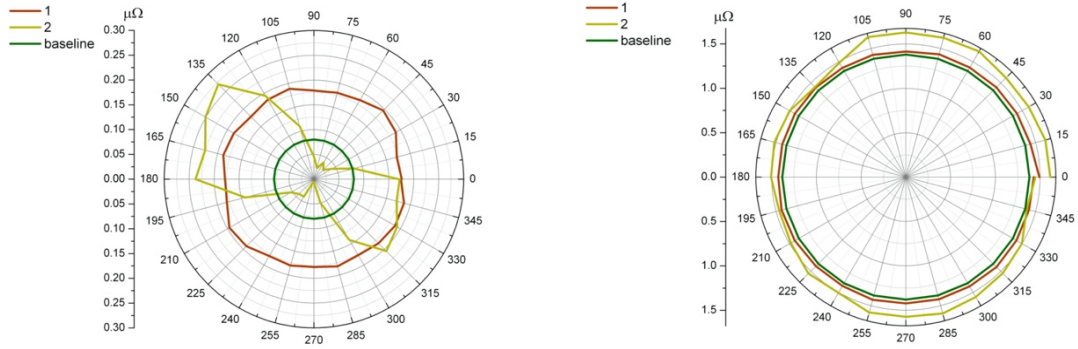

Figure B2, Transverse impedance of the rotational TME map along the y-axis @ 10 kOe and 4.2 K at 2 > ν > 1 in the fourth test. The sample z-axis is vertical at 0°, which is parallel to the gravity and the applied *H* field. Baselines are the reference lines taken at 10 K before applying the *H* field, and are shifted by arbitrary levels to present the impedance of switching sign better. Line 1 is taken at the *H*-up phase while line 2 is taken at the *H*-down phase. *Left*) Resistive part. There are pronounced negative contributions at 90° and 270°, where the sample y-axis is in the vertical direction. *Right*) Inductive part. There are pronounced geometric phases, i.e., the transverse inductances are distinct after 360° rotation. The difference between 0° and 360° became acute at $H \geq$ 20 kOe, i.e., the 360° inductances of both line 1 and line 2 were negative (not shown here).

In the fourth test, we initially placed the sample at $\theta_R$ = 0°, i.e., the sample z-axis was parallel to gravity and the applied *H* field. The condition at $\theta_R$ = 0° is almost the same as that in the second and third tests, except that the residual *H* and the earth's field are slightly different in Beijing. The condition at $\theta_R$ = 90° is almost the same as that in the first test, i.e., the y-axis is vertical while the x-axis is along the N-S direction, except that the residual *H* in the dewar is much higher than the earth's field.

We took the rotational results at 10 K before applying *H* to be the zero-TME reference lines, which are the baselines in Fig. B2. The sample returned to 0°, whenever we changed the *H* settings. Line 1 is in the *H*-up phase, i.e., raising *H* from 0 to 90 kOe at 4.2K. Line 2 is in the *H*-down phase, i.e., lowering *H* from 90 to 0 kOe at 4.2K. The sample wiring caused a small inductance, which showed a constant amplitude regarding *T*, *H*, $\theta_R$, but not $f_d$. We shifted the baselines to higher levels in Fig. B2 to present the negative contributions better. For example, line 2 has negative resistances near 90° and 270°. Both line 1 and line 2 have negative inductances at $\theta_R$ = 360° for $H \geq$ 20 kOe, as reported previously [11]. However, both of them are still positive at 10 kOe in Fig. B2. The RRP vortices rotated by the *H* field in the sample are pinned by boojums, and decay extra slowly because of the low thermal detachment at 4.2 K.

We now reassess Figure 2 in reference [11] to reveal some important effects, because they are more transparent at 90 kOe than that at 10 kOe. The $H$ field drags the RRP vortices to rotate along with the $\theta_R$ settings. The length and the twist of the RRP vortices readjust themselves abruptly, when rotating from the short z-axis to the long y-axis, and this causes a zig-zag TME response whenever $\theta_R$ changes. We carried out the clockwise and counter-clockwise rotations twice, which show different results for the same $\theta_R$ settings at each rotation no matter whether they are forwards or backwards. The TME results retain some memory even when the rotation returns from $\theta_R$ to 0°. The transverse inductances change sign from positive to negative after the 360° rotation, which return to positive by rotating back to 0° every time. The transverse open-loop responses grow with $H_z^2$. We thus suggest that the loosely overlapping γ-dopants move separately on a $\nu = 1$ vacuum, which represents the filled AF ladders along the x-axis. The high $H$ field causes some RRP vortices to incline out of the x-y plane, which are terminated by the emergent boojums to perform an anisotropic STB oscillation. No matter what the reason is for the transverse inductance, the $H$ field carries the inclined RRP vortices to circle the Dirac nodes over a $2\pi$ angle such that the particle-like vortices become the hole-like anti-vortices. More precisely, the change in vertical RRP $l_z$ from 1/2 to -1/2 causes a geometric π phase in the STB mode.

Figure B2 shows the hysteresis of the rotational TME map at 10 kOe, which reveals several important facts. The rotation detaches some RRP vortices, which recover AF ladders with a lower $n_L$ when lowering $H$ from 90 to 10 kOe. Line 1 shows only one kind of RRP vortex in the sample except for some possible anapoles because of ν > 1, which are perpendicular to the $H$ field. Line 2 shows more than one kind of RRP vortex. Some high-$n_L$ RRP vortices remain due to the slow thermal mixing [67] when lowering $H$ stepwise from 90 to 10 kOe, which are parallel to the $H$ field. The line-2 impedance at 90° and 270° reveals the chiral anomaly of $E_y$. The negative contributions of the line-2 resistance near 90° and 270° are caused by extending the RRP length from 1.2 to 10 mm, which increases the STB amplitude. The high-$n_L$ RRP vortices in line 2 provide a stronger coupling with the cosmic axion field. Besides this, the longitudinal resistance in line 2 also carries a negative sign associated with the gain in energy at 90° and 270° (not shown). Rotating the RRP vortices between the long and short axes causes a canted rotational map of the line-2 resistance.

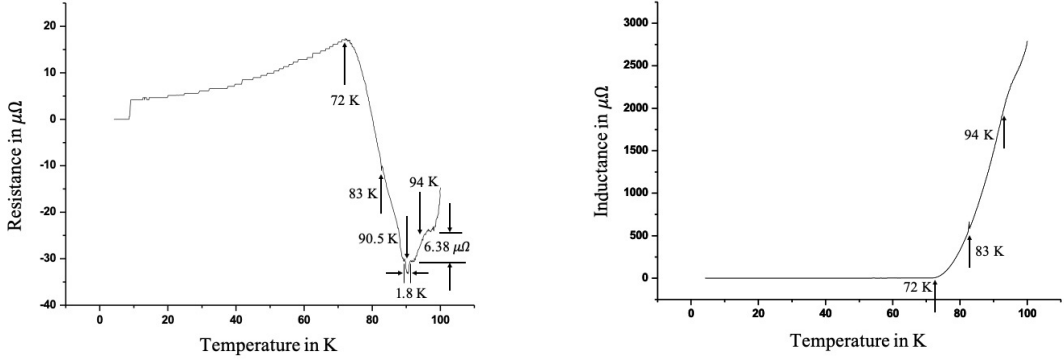

Figure B3, TME impedance of the anomalous topological Kondo effect at the residual $H_z \sim 60$ Oe in the dewar. The measurement started at 100 K, which slowly cooled down to 4.2 K. We turned the drive current on with 13 mA and 4 Hz, before we lowered the $H_z$ directly from 90 to 0 kOe at 100 K. The resistive part contains rich structures, whereas the inductive part is quite smooth except for a common spike at 83 K. *Left*) Resistive part. The negative resistivity of the anapoles is located between 72 and 100 K. The resistance between two plateaux is 6.38 μΩ. The sag at 90.5 K has a width of 1.8 K. *Right*) Inductive part, which persisted in the presence of the anapole at $T > 72$ K.

*B.3 Anomalous topological Kondo effect in the second and fourth tests*

We observed the anomalous topological Kondo effect of $T_{Ka} \sim 61$ K by lowering $T$ from 100 to 4.2 K in the day-four study of the second test (see Fig. B3). More precisely, the results of cooling from 100 to 4.2 K were obtained after lowering $H_z$ rapidly from 90 to 0 kOe at 100 K. More than 80% of the $H_z^2$-magnets ~ 0.13 mH at 90 kOe in Fig. 4 was preserved. Had there been no Floquet-Kondo screening at all, the whole 0.13 mH would have been retained at the residual $H_z \sim 60$ Oe.

The Korringa constant is a coefficient used to achieve mutual thermal equilibrium between the nuclear and electronic spins [67], where the interactions among nuclei are much weaker than that of the metallic electrons. This relationship is reversed in our case, i.e., the coherent nuclear interactions are much stronger than those of the Mott electrons. The equalization time between two heat reservoirs is exceedingly long when lowering $H_z$ rapidly, which is attributed to an extremely large Korringa-like constant because of the $N_I$ collectivity of nuclei. The coherent nuclei almost decouple Mott electrons without applying $H_z$ at $T > T_{Ka}$.

We previously reported a similar case [11], when $T_{Kl}$ was ~ 225 K at $\nu > 1$ in the fourth test. The RRP fluxes were rotated to the shortest z-axis by applying $H_z = 90$ kOe at 400 K, which survived 300 K for some days at the earth's field, as visualized using magnetic powders [11]. The high-$n_L$ lottons are stretched by boojums. Twisted and stretched RRP vortices decay to AF ladders mainly by means of the thermal detachment, where the γ radiation and the "light shining through walls" [80] are irrelevant. Figure B2 reveals another similar case at $\nu > 1$ in the fourth test but at 4.2 K, where some high-$n_L$ RRP vortices are preserved by lowering $H_z$ step-by-step from 90 to 10 kOe.

The physics become very different in the presence of the Floquet drive, when we remove the applied $H_z$ completely and rapidly. Before the RRP vortices transfer to AF ladders on the x-y plane, the annular drive field catches the detached RRP vortices to construct the AF and gofour anapoles. Now, the ring-type anapole recovers the spontaneous $B_L$ ~17 mT. The flavour switches from RRP vortices to the condensed anapole superfluid due to the high packing factor $\nu > \nu_{qc}$.

Figure B3 illustrates the complicated TME of the anapole condensate in the cooling process. The inductance decreased linearly from 0.1 to 0 mH between 100 and 72 K, where the phase $> \pi/2$ at 100 K decreased to a zero phase at 72 K. The resistance showed an S-shaped swing with a peak-to-peak amplitude ~ 50 μΩ superposing the normal resistance ~ 20 μΩ, which went down and then up during the cooling process from 100 to 72 K. More precisely, the absolute negative resistance launched a sag ~ -30 μΩ at 90.5 K. The sag width was 1.8 ± 0.2 K.

We turned the drive current off after $T$ reached 4.2 K in Fig. B3. The complicated $T$-behaviours disappeared in the absence of anapoles, as revealed by a calibration process using 10 Hz and raising $T$ from 4.2 to 81 K later that night. Regardless of whether the driving was either $T$-up or $T$-down in the second test, the resistance showed always a smooth and monotone $T$-function between 10 and 300 K irrespective of the frequency used, apart from the small slope of the swing that was barely identifiable near $T_{Kl}$.

We carefully calibrated the phase by demanding a purely resistive transition of the superconductor at 9.3 K (see Fig. B3). The resistance of 6.38 μΩ = $8\nu R_K/N_I$ between two $T$-plateaux reveals the octonion nature, which also implies an irrelevant systematic phase error < 0.02 mRad near 90 K. The smooth inductive curve further supports this precise calibration over the whole $T$-range, where the rich structures of resistance vanish except for a common spike at 83 K. The $T_{Kl}$ estimate using $f_{LLL}$ also confirms the phase calibration. Hence, we validate the absolute negative resistance to gaining energy.

We provide an essentially phenomenological description in the following section, but before that we apply some universal considerations to validate the anapole condensate. Anapoles have an annular nuclear polarization at $T > T_{Ka}$, where the Mott electrons no longer screen the nuclear magnets. Had we kept $T$ at 100 K, the 0.1-mH induction would have been preserved for days because of the extra long equalization time. Hence, we lost the 20% induction at the moment of the flavour transition, as caused by the dynamical g-wave screening at 100 K.

We assume a constant $n_L$ for several minutes to cool $T$ from 100 to 72 K. The continuous inductance $\propto T$ implies a thermal equilibrium in the heat reservoir of Mott electrons. It is symmetrical, no matter how we drive the $T$ up or down between 72 and 100 K, as long as $T$ is changed much more slowly than the electronic relaxation time. The magnetic polarization decreases linearly with $(T - T_{on2})$ with $T_{on2}$ ~ 72 K, where Mott electrons screen the ring-type Dirac strings of nuclear magnets more and more. The linear thermalization $\propto T$ implies gaps among magnetic quanta with a magnitude $< k_B T_{on2}$.

The quantum anomalous $T$-plateau implies an incompressible quantum fluid to conserve the flavour symmetry, which are caused by the commensuration in (7). The negative resistive contribution reveals the anapole condensate at least between 72 and 100 K.

Figure B2 shows two turning points and three sequential *T*-points spanned by 11 K, i.e., 72, 83, 94 K. The 4-Hz Floquet drive splits the anapole minigap levels with a time-varying $B_L^2$ term, which become Kondo doublets separated by $2hN_I f_d = k_B \times 11.0$ K in energy. Three *T*-points are thus $T_{Ka} + 2mhN_I f_d/k_B$, where *m* is the magnetic quantum number of Mott electrons. We thus read $T_{Ka} \approx 61$ K according to this model.

The first turning point is triggered by the sag on the *T*-plateau at 90.5 K. The common spike at $T_{Ka} + 4hN_I f_d/k_B = 83$ K indicates that the overlap of d-wave Kondo doublets vanishes, i.e., the screening of d-wave Mott electrons vanishes at *T* > 83 K. The second turning point is located at $T_{on2} = T_{Ka} + 2hN_I f_d/k_B$, where the overlap between p-wave Kondo doublets vanishes. Hence, the induction of ion magnets vanishes completely at *T* < $T_{on2}$. We thus speculate that the inductive slope change at $T_{Ka} + 6hN_I f_d/k_B = 94$ K is caused by a minor f-wave effect. Once anapoles were caught by the Floquet drive, it always locked and loaded during the cooling process until we turned the Floquet drive off.

The stepwise continuous resistive function reveals that the clockwise and counter-clockwise anapoles are paired by small gaps, i.e., $\Delta_a$ and $\Delta_{AF}$. Two *T*-plateaux separated by $8\nu R_K/N_I$ reveal the octonion nature of gapped anapoles.

The anomalous topological Kondo effect became acute after the sample reactivation with $\nu > 1$ one year later. The Kondo anomaly appeared at ~ 225 K during a cooling process from 300 to 4.2 K (see Appendix B2), which depended slightly on $f_d$ and $I_x$. The longitudinal impedance jumped abruptly up to several levels on the mΩ order at *T* ~ 225 K, which depended significantly on $f_d$ and even $I_x$. A small transverse resistance decreased smoothly with *T*, and was independent of $f_d$ and $I_x$. This remarkable observation suggests a ring-like structure, i.e., only the longitudinal current threading the anapole-like excitations yields the TME response. However, the possible anapole has not yet been studied that well.

We also monitored the anomaly in real time. The amplitude of the ac response went up slowly and was accompanied by a phase shift toward the inductive side when quenching the sample in LN$_2$. It returned to the normal resistance after we took it out. The stationary impedance was quite unstable in LN$_2$, which depended significantly on the drive parameters, as mentioned above.

*B.4 GW signal in the second test*

We copied two pictures taken on June 26$^{th}$ 2011 from the experimental note book. Figure B4 shows the amplitude of GW signals, where the phase information is absent. The calibration reveals a capacitive bias of ~ -13*i* nV using $f_d$ ~ 5 Hz. GWs flip the pseudospin of the seesaw quadrupole edge states one-by-one as sequential events, which give rise to the sawtooth modulations characterized by a single $f_{GW}$. The TME behaves like a capacitor, which yields a smaller sawtooth amplitude by a higher $f_{GW}$. However, each GW signal has a constant sawtooth amplitude, which is independent of the GW strain amplitude arriving at earth. We scanned $f_d$ to select one of the GWs using the matching condition $|f_d - f_0| < 1/\tau_{R2}$. The exact frequency $f_0 = 4 \times 92 f_{LLL} + 2 f_{LLL}$ is very close to 4.8903 Hz, as explained in the main text.

The left panel of Fig. B4 shows two distinct GW signals ~ 0.5 mHz at noon, which are acquired by $f_d$ = 4.890, 4.8895, and 4.891 Hz, respectively. We resolved the first GW signal better at 0.53 mHz while we resolved the second GW signal poorly at ~ 0.48 mHz. Two GW signals constrain $f_0$ between 4.890 and 4.891 Hz, because the signal propagates backwards, when $f_d$ = 4.8895 < $f_0$ changes to $f_d$ = 4.891 > $f_0$.

The matching condition picks up one and only one GW signal every time. For example, the same $f_d$ = 4.893 Hz picked up two distinct GWs of 1.5 and 3.6 mHz. The earth's rotation switched the selection from 1.5 to 3.6 mHz. However, the extinguished GW signal of 1.5 mHz returned, when we lowered $f_d$ from 4.898 to 4.888 Hz. According to this observation, we adopt $f_0$ = 4.8903 Hz ± 0.2 mHz. Note that the signal propagates backwards again, when $f_d$ = 4.898 > $f_0$ changes to $f_d$ = 4.888 < $f_0$.

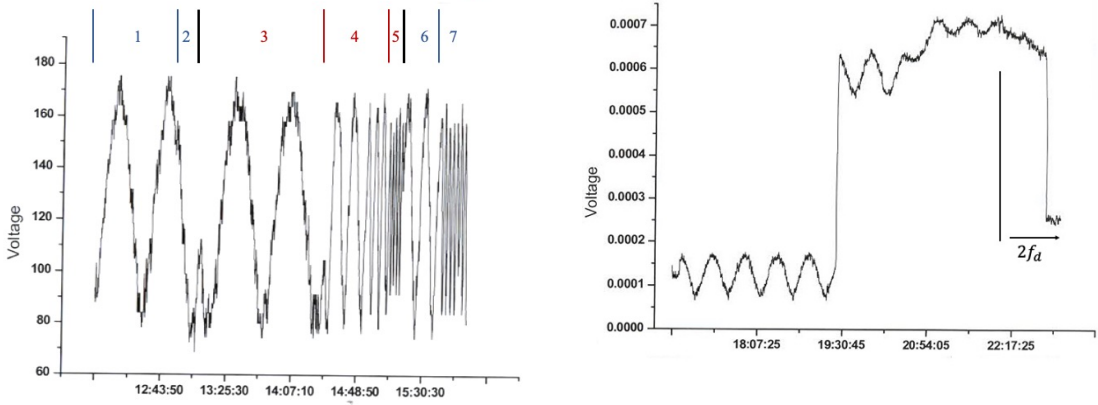

Figure B4, Evolutions of the GW signals in the second test in Hsinchu. *Left*) Constant temperature at 4.2 K. The ordinate unit is 0.1 nV. We scanned $f_d$ using (1, 4.89 Hz < $f_0$), (2, 4.8895 Hz < $f_0$), (3, 4.891 Hz > $f_0$), (4, 4.893 Hz > $f_0$), (5, 4.898 Hz > $f_0$), (6, 4.888 Hz < $f_0$), and (7, 4.883 Hz < $f_0$) Hz, where the red index number shows $f_d > f_0$, the blue index number shows $f_d < f_0$, and $f_0 \cong 4.8903$ Hz. *Right*) Switching the temperature between 4.2 and 10 K using $f_d$ = 4.889 Hz and $2f_d$ = 9.7792 Hz. The residual resistance of 4.5μΩ sent the signal to a higher level and created a backward propagation of the 0.53-mHz signal. The GW signal persisted in the absence of superconductivity. The ordinate unit is 0.1 mV. We changed the drive frequency from $f_d$ to $2f_d$ near 10 pm.

The right panel of Fig. B4 shows the 0.53-mHz signal using $f_d$ = 4.889 Hz, which was independent of *T*, and more particularly independent of superconductivity. However, the GW signal evolved backwards in time at the moment of losing superconductivity. The signal at 0.53 mHz faded out gradually in the absence of superconductivity. We changed the frequency from $f_d$ to 9.7792 Hz before restoring the superconductivity near 10 pm, because we guessed $f_0$ ~ 4.8896 Hz at that time. The systematic bias became ~ -25*i* nV using 9.7792 Hz. It is just possible to see a very slow signal in Fig. B4 using the $2f_0$ drive at 10 K, which is probably not a GW signal. In contrast, the 2.448 Hz drive yielded a zero TME at 4.2 K.